\documentclass{aa} 
\usepackage[english] {babel}
\usepackage{astron}
\usepackage[dvips]{graphicx}
\usepackage{ae}
\usepackage{color}
\usepackage{amsfonts}
\usepackage{tabularx}
\usepackage{natbib}
\usepackage{amssymb,amsmath}
\usepackage{listliketab}
\usepackage[toc,page]{appendix}
\usepackage{txfonts}
\usepackage{placeins}
\usepackage{float}
\usepackage{afterpage}
\usepackage{longtable}
\usepackage{footmisc}
\usepackage{supertabular}
\usepackage{url}
\usepackage{rotating}
\usepackage{wasysym}
\usepackage{adjustbox,lipsum}
\usepackage{pdflscape}
\usepackage[breaklinks]{hyperref}
\usepackage[hyphenbreaks]{breakurl}
\usepackage{supertabular}
\begin{document}
\bibliographystyle{aa} 

   \title{Hubble flow variations as a test for inhomogeneous cosmology\thanks{Tables \ref{sample_cat_fi_lg} and \ref{sample_cat_fi_cmb} are only available in electronic form at the CDS via anonymous ftp to cdsarc.u-strasbg.fr (130.79.128.5) or via \url{http://cdsweb.u-strasbg.fr/cgi-bin/qcat?J/A+A/}}}
   
   \author{Christoph Saulder
          \inst{1,2}
          \and
          Steffen Mieske
          \inst{3}
          \and
          Eelco van Kampen
          \inst{4}
          \and
          Werner W. Zeilinger
          \inst{2}                
          }

   \institute{
   Korea Institute for Advanced Study,
   85 Hoegi-ro, Dongdaemun-gu, Seoul, South Korea\\
   \email{christoph.saulder@equinoxomega.net}\\
   \and
   Department of Astrophysics, University of Vienna,
   T\"urkenschanzstra\ss e 17, 1180 Vienna, Austria\\
       \and   
   European Southern Observatory,
   Alonso de C\'{o}rdova 3107, Vitacura, Casilla 19001, Santiago, Chile\\
       \and
   European Southern Observatory,
   Karl-Schwarzschild-Stra\ss e 2, 85748 Garching bei M\"unchen, Germany\\
}

   \date{Received June 23 2016 ; accepted November 28, 2018}

\abstract
{
Backreactions from large-scale inhomogeneities may provide an elegant explanation for the observed accelerated expansion of the universe without the need to introduce dark energy.}
{
We propose a cosmological test for a specific model of inhomogeneous cosmology, called timescape cosmology. Using large-scale galaxy surveys such as SDSS and 2MRS, we test the variation of expansion expected in the $\Lambda$-CDM model versus a more generic differential expansion using our own calibrations of bounds suggested by timescape cosmology.
}
{
Our test measures the systematic variations of the Hubble flow towards distant galaxies groups as a function of the matter distribution in the lines of sight to those galaxy groups. We compare the observed systematic variation of the Hubble flow to mock catalogues from the Millennium Simulation in the case of the $\Lambda$-CDM model, and a deformed version of the same simulation that exhibits more pronounced differential expansion.
}
{
We perform a series of statistical tests, ranging from linear regressions to Kolmogorov-Smirnov tests, on the obtained data. They consistently yield results preferring $\Lambda$-CDM cosmology over our approximated model of timescape cosmology. 
}
{
Our analysis of observational data shows no evidence that the variation of expansion differs from that of the standard $\Lambda$-CDM model.
} 


   \maketitle

\section{Introduction}
According to the $\Lambda$-CDM model, the universe consists of about 68\% dark energy, 27\% dark matter, and 5\% baryonic matter and it is about 13.8 Gyr old \citep{Planck}. This model provides a widely accepted and successful description of the general behaviour and appearance of our universe. But there is a major problem: more than 95\% of the total energy content of the universe is hidden from direct observations. The nature of dark matter is still an enigma, while there are some observations that appear to provide direct empirical evidence for its existence \citep{BulletCluster}. Dark energy, which makes up more than two-thirds of the Universe's total energy content, is  the greatest mystery in cosmology today. For more than a decade its true nature has puzzled physicists and astronomers alike. There have been numerous attempts \citep{Zlatev:1999,Steinhardt:1999,Armendariz:2000,Kai:2007,Mavromatos:2007,Alexander:2009} to explain this phenomenon. The current model, a cosmological constant that adds an additional term to the Einstein equations of general relativity provides a relatively simple way to fit the observed data. An oddity is that the value of the cosmological constant derived from quantum fields and standard model particle physics (Higgs condensate) is about $10^{56}$ times larger than the value actually observed \citep{Bass:2011}. Most attempts to explain the accelerated expansion of the Universe require either new physics ($\Lambda$ as an extension of `classic' general relativity) or some special matter distribution \citep{Zibin:2008}. 

In an alternative approach, it is also possible to take one step back to the very basics of modern cosmology. The cosmological principle states that the Universe is homogeneous and isotropic. However, this is not true on all scales: the Universe is made of galaxies, clusters, and voids and not a homogeneous distribution of stars, gas, and dark matter. Only when one reaches scales of at least 100-150 Mpc \citep{Hogg:2005}, can one average all smaller structures and the cosmological principle is fulfilled. The cosmological principle itself is very useful, because in the case of homogeneity and isotropy, one is able to find a simple solution of Einstein's field equations of general relativity: the Friedmann-Lema\^{i}tre-Robertson-Walker metric. Dropping the assumption of homogeneity leads to different solutions, in particular timescape cosmology \citep{Wiltshire:2007}. 

The majority of research that has been carried out on inhomogeneous cosmology so far has been theoretical in nature. The general idea of taking inhomogeneities into account in cosmology is already quite old and most often studied using exact solutions of the Einstein equations with a pressure-less dust source\footnote{In the context of general relativity, this means an energy-momentum tensor $T^{\mu \nu}= \rho U^{\mu} U^{\nu}$ composed of the matter density $\rho$ and the four-velocity $U^{\mu}$.} \citep{Lemaitre:1933,Tolman:1934,Bondi:1947,Szekeres:1975}, or inhomogeneities locally inserted into a spatially homogeneous Universe with metric junction conditions \citep{Einstein:1945,Einstein:1946}. Some notable considerations on the subject of more general inhomogeneities were undertaken by \citet{Ellis:1984},\citet{Ellis:1987}, \citet{Ellis:1989}, \citet{Zalaletdinov:1992}, and \citet{Harwit:1995}. Substantial progress has been made in the description of the effects of inhomogeneities in the context of general relativity in the past two decades. The influence of inhomogeneities on the average properties of cosmological parameters has ben considered in several works \citep{Buchert:2000a,Buchert:2000b,Buchert:2000,Buchert:2001} using both perturbation theory and in full general relativity. Since Einstein's field equations are a set of ten non-linear partial differential equations, one cannot average as usual (in the case of linear equation) if there are significant inhomogeneities (such as very empty voids and clusters with densities far higher than the critical density of the Universe). A backreaction (feedback) caused by these inhomogeneities is expected due to the non-linear nature of general relativity. This backreaction and a recalibration of average spatial volumes in the presence of spatial curvature gradients cause the observed (or `dressed') values of cosmological parameters to be different from the `real'(or bare) values \citep{Buchert:2003}. Therefore, one has to recalibrate cosmological measurements, which were made under the assumption of a homogeneous Universe (Friedmann equations), in the framework of inhomogeneous cosmology. In the simple case of general relativistic dust, the so-called Buchert's scheme \citep{Buchert:2000}, a set of modified Friedman equations is obtained. A few years later it was shown that the acceleration expansion of the Universe cannot be fully understood in a simple pertubative approach alone \citep{Rasanen:2006,Kolb:2006,Ishibashi:2006}. 

One of the most advanced models of an inhomogeneous cosmology, which can mimic dark energy, was presented in \citet{Wiltshire:2007} and it is called `timescape cosmology'. It uses a simple two-phase model (with a fractal bubble or Swiss cheese like distribution of matter) consisting of almost empty voids (very low density regions devoid of any galaxies) and dense walls (clusters and filaments). The concept of finite infinity \citep{Ellis:1984,Wiltshire:2007} is introduced, which marks the boundary between regions that may become gravitationally bound and regions that are expanding freely due to the Hubble flow. In the timescape model one has to treat both areas independently. Inside a finite infinity boundary, the average geometry can be approximated to be flat and have an average matter density corresponding to the re-normalized critical density of timescape cosmology. Voids, however, are defined by an open geometry and are of extremely low density (close to empty). In this model, aside from homogeneity, the assumption of a universal cosmological time parameter is also dropped. Hence, the backreactions from inhomogeneities cause significant differences in the time flow, due to effects of quasilocal gravitational energy, so that the Universe in the middle of a void is older than in the centre of a cluster by several Gigayears (hence the name timescape cosmology). 
For a wall observer like ourselves, the voids would appear to expand faster than the walls. In a very simplified picture, this can be imagined such that at large scales these different expansion rates are manifested in an apparent accelerated expansion of the Universe for an observer located in a wall-environment, because the fraction of the total volume in the Universe occupied by voids constantly increases due to their higher expansion\footnote{From the point of view of a wall observer, because of the different clock rates in both regions.} rate and structure formation. Consequently, the average expansion rate approaches the void expansion rate in later times. A much more detailed mathematical description of the model and the derivations of its properties can be found in \citet{Wiltshire:2007,Wiltshire:2009,Wiltshire:2013}.

Timescape cosmology and similar inhomogeneous cosmologies might be possible solutions for the dark energy problem, because they can qualitatively predict a signal, which could misinterpreted as an accelerated expansion of a Universe with a Friedmann-Lema\^{i}tre-Robertson-Walker metric. However, estimates of the magnitude of the backreactions from inhomogeneities and their influence on the expansion of the Universe are difficult (due to the non-linearity and complexity of the equations) and range from negligible to extremely important \citep{Marra:2010,Mattsson:2010,Kwan:2009,Clarkson:2009,Paranjape:2009,vdHoogen:2010,Smale:2011}. Lately, new arguments appeared that if backreactions are not sufficiently strong effect to fully explain the signal attributed to dark energy, they will still have an impact on the cosmological parameters and distance measurements on a few percent level \citep{Clarkson:2012,Umeh:2014a,Umeh:2014b,Clarkson:2014}. In contrast to these claims, the calculations of \citet{Kaiser:2015} suggest that such an effect would be insignificantly tiny, which is supported by other recent work \citep{Lavinto:2015}. The issue of the scale and relevance of backreactions remains an ongoing discussion \citep{Buchert:2015,Kaiser:2017,Buchert:2018}. This shows that observational tests for timescape cosmology are essential for the ongoing debate and may also help to better understand similar models. Several tests for timescape cosmology were proposed in \citet{Wiltshire:2010,Wiltshire:2011,Wiltshire:2013}, most of which are rather complex. So far, they have not been able to produce striking evidence neither for nor against timescape cosmology, and timescape cosmology remains within the uncertainties of current observational data \citep{Smale:2011,Duley:2013,Sapone:2014,Nazer:2015,Dam:2017}. 

In this paper, we focus on a conceptually simple experiment for differential expansion based on the timescape cosmology \citep{Wiltshire:2007}. This is done by comparing observational data to predictions from the standard model ($\Lambda$-CDM cosmology) and this alternative theory. Both theories explain the observed accelerated expansion \citep{Riess:1998,Perlmutter:1999,Schmidt:1998} of the Universe, however in two radically different ways. 
\section{Concept of the cosmological test}
\label{sec_test}
The idea of the test is to discriminate between the two theories, $\Lambda$-CDM cosmology and timescape cosmology, by looking for a specific signal in the observational data, which is predicted in one theory, but has a significantly different magnitude in the other. Differential expansion is a feature expected even in the most basic inhomogeneous cosmological models \citep{Tolman:1934,Szekeres:1975}. Timescape cosmology predicts that even though the expansion of the Universe occurs at a different rate in cosmological voids as compared to cosmological walls, when referred to the clocks of any one observer, the clock rates of canonical observers also vary so as to keep the quasilocally measured expansion uniform on large scales. This is on account of gravitational energy gradients arising from spatial curvature gradients. Conceptually, this follows as a consequence of the Cosmological Equivalence Principle proposed by \citet{Wiltshire:2008}. Using the differential expansions of cosmological voids and walls as an observational feature would present an interesting test for timescape cosmology \citep{Schwarz:2010}. The main goal of this work is to see if the expansion rates for voids and walls are systematically different in the way predicted by timescape cosmology or if they are in agreement with the predictions of $\Lambda$-CDM cosmology. In this work our goal is to directly measure these different expansion rates. 

The initial concept of our test was outlined in our previous work \citep{Saulder:2012}. A significantly improved version is presented in this paper, because to provide a solid test, one has to consider and measure all possible biases and calibrate the tools required for the test very carefully to minimize systematic effects. One of these biases that is present in any reasonable cosmology and which is discussed in Section \ref{sec_infall} in more detail, is the coherent infall of galaxies into large structures that cause a similar, though likely smaller signal than a generic differential expansion.

\begin{figure*}[ht]
\begin{center}
\includegraphics[width=0.90\textwidth]{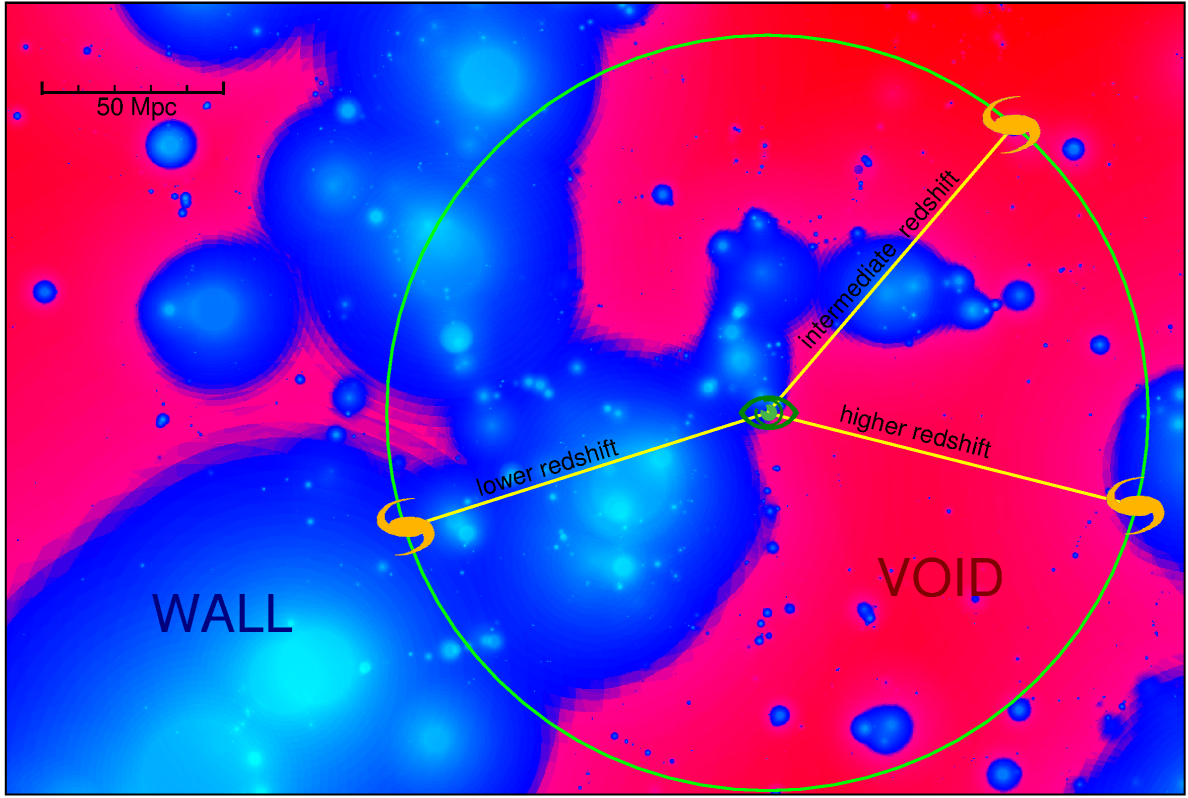}\\ 
\caption{Sketch illustrating the line of sight dependence of the individual Hubble parameter. One of the expectations of timescape cosmology is that a galaxy (yellow galaxy symbol) at the certain distance from a observer (green eye symbol) may have a higher redshift with primarily void environment (reddish areas) in the line of sight than another galaxy at the exactly same distance but with primarily wall environment (bluish areas) in the line of sight.}
\label{scetch_new}
\end{center}
\end{figure*}

One of the most notable (compared to $\Lambda$-CDM cosmology) feature of timescape cosmology is that voids expand faster than walls and that theory provides clear constraints on their values to be compatible with other cosmological observations. The difference in the expansion rate should be measurable at cosmologically small scales (a few hundred Mpc, \citet{Schwarz:2010}). If one observes a galaxy by looking through a void, its redshift is expected to be greater than that of another galaxy at the same distance observed along a wall in the framework of timescape cosmology. Assuming that timescape cosmology is a valid description of the Universe, the (bare, in the language of timescape cosmology) Hubble parameter for dense environment is expected to be 50.1 $\pm$ 1.7 km $\textrm{s}^{-1}$ $\textrm{Mpc}^{-1}$ and the average (from inside a wall) observed (dressed, in the language of timescape cosmology) Hubble parameter to be 61.7 $\pm$ 3.0 km $\textrm{s}^{-1}$ $ \textrm{Mpc}^{-1}$ \citep{Duley:2013} based on the at that time current Planck results \citep{Planck,Planck_cosmo_old}. Very similar values were found earlier by \citet{Wiltshire:2007,Leith:2008} according to the best fit on supernovae Type Ia \citep{Riess:2007}, cosmic microwave background (CMB) \citep{WMAP_pre,WMAP3} and Baryonic acoustic oscillations \citep{Cole:2005,Eisenstein:2005} data and later by \citet{Nazer:2015}, albeit with larger error bars. Adopting those values, timescape cosmology can reproduce the observed accelerated expansion without having to introduce dark energy. The measured Hubble parameter depends on the density profile of the line of sight to a galaxy (see Figure \ref{scetch_new}). Since voids make up the largest volume fraction of the Universe now, the value of the dressed Hubble parameter should be closer to the one of the void environment. For the same reason, it will be easier to find galaxies with many voids in their lines of sight. Galaxies with mainly a high density environment in line of sight, which means the larger part of the line of sight is located within finite infinity regions, will be rarer. However, the average value of the Hubble parameter in timescape cosmology is still lower than the usual values for $\Lambda$-CDM cosmology \citep{Planck_cosmo} due to inherent differences between both cosmologies. In timescape cosmology,the lapse of time between average observers in galaxies within finite infinity regions and those in negatively-curved voids is different \citep{Wiltshire:2007}. The latter would observe a cooler CMB and a much older Universe. Fitting our own observations to light rays that traverse both finite infinity and void regions on cosmological scales results in a lower average Hubble constant and a slightly older Universe than standard. To test the validity of timescape cosmology using the differential expansion rates, one has to compare the redshift of a galaxy to another independent distance indicator or standard candle to calculate its `individual Hubble parameter' (the Hubble parameter measured for one singular galaxy or cluster) $H_{\textrm{i}}$. Since the area of interest for this investigation ranges up to several 100 Mpc, it cannot be covered by Cepheid variable stars with present observing tools and supernovae type Ia are too rare. Consequently, one has to use techniques like the surface brightness fluctuation method, the fundamental plane of elliptical galaxies, the Tully-Fischer relation or similar methods. We choose the fundamental plane of elliptical galaxies \citep{Dressler:1987,Djorgovski:1987}, because it can be applied to sufficiently large cosmological distances and be best realised with automatic pipelines and large datasets. 

Furthermore, a solid model of the matter distribution in the local Universe is required to derive the finite infinity regions. Our datasets do not have to be extremely deep, because the variations in the Hubble flow predicted by timescape cosmology can only be measured below the scale of homogeneity, which is of the order of $\sim150$ Mpc. Beyond that distance, the ratio of the void and wall environment in the line of sight approaches a constant value. 

While the basic idea of this test is simple, the test needs to be executed with great care to correctly take into account any relevant systematic effects and biases. This includes effects, which are expected in any cosmology, such as the coherent infall of galaxies into clusters, for which we can use mock catalogues based on cosmological simulations. Additionally, one has to consider that by constructing the finite infinity regions from galaxies (bottom-up), we are extrapolating the ideas of the current theoretical construct (top-down) of timescape cosmology, beyond the well-studied (theoretical) regime. 

\section{Data}
Most of the data which we used for the cosmological test was already prepared in our previous work. As mentioned before, we require a redshift independent distance indicator which can be applied to a large number of galaxies in the local Universe. To this end, we used the fundamental plane of elliptical galaxies, which we calibrated and discussed in detail in \citet{Saulder:2013}. We provided slightly improved coefficients in the Appendix of \citet{Saulder:2015a}. We used Malmquist-bias corrected least-square fits to obtain the coefficients from SDSS data \citep{SDSS_DR10}, in which we identified early-type galaxies using GalaxyZoo \citep{GalaxyZoo,GalaxyZoo_data}. In \citet{Saulder:2016}, we combined our fundamental plane data with the SDSS based group catalogue of that paper. This further improved the fundamental plane based distance estimates for galaxy groups that host more than one early-type galaxy. The fundamental plane distance catalogue of \citet{Saulder:2016} provides one part of the required data to test timescape cosmology.

In \citet{Saulder:2016}, we also provided a catalogue of finite infinity regions in the local Universe. We derived the finite infinity regions from the masses of SDSS DR12 \citep{SDSS_DR12} and 2MRS \citep{2MRS} galaxy groups, which were published in the same paper. However, we used a slightly updated version of that catalogue, which is provided with this paper (see Appendix \ref{newfiregions}). The catalogue of finite infinity regions completed the observational data required for our test. The finite infinity regions were calculated assuming timescape cosmology. While one can define the concept of finite infinity regions also in $\Lambda$-CDM cosmology, we intentionally only used the timescape definition for consistency reasons, because we aimed to observe an effect predicted by timescape cosmology. 

We also needed simulated data to compare the theoretical predictions to the observations. To this end, we used the Millennium simulation  \citep{millennium,Mill_data} and the semi-analytical galaxy models from \citet{Guo:2011} implemented in the Millennium simulation results. In \citet{Saulder:2016}, we had already used this data to create mock catalogues to calibrate our group finder algorithm. Here, we used these mock catalogues again and in combination with some further calibrations. We also required some additional data from the Millennium simulation to identify early-type galaxies. With this data and some additional considerations, which are discussed in the next section, we were able to create mock datasets for the expectations of $\Lambda$-CDM cosmology and an approximation of timescape cosmology obtained by deforming a $\Lambda$-CDM model. 

\section{Method}
\label{sec_method}
We used the groundwork laid in \citet{Saulder:2013} and \citet{Saulder:2016} to perform our test of timescape cosmology on the observational data. We took the fundamental plane distances and redshifts from the fundamental plane distance catalogue of \citet{Saulder:2016} to calculate the relative individual Hubble parameters for all groups in SDSS hosting early-type galaxies with a redshift lower than 0.1. The relative individual Hubble parameters for the observational data were obtained by dividing the group's redshift by its co-moving fundamental plane distance and normalizing the results in a way that the averaged Hubble parameter of the sample corresponds to unity. The finite infinity regions which were also published in the finite infinity regions catalogues of \citet{Saulder:2016}, were approximated by spherical regions with an on average critical density (of the timescape model). While the before-mentioned catalogues and dataset were all originally calculated in the rest-frame of the CMB, we additionally calibrated the entire data for the rest-frame of the local group, because the CMB-frame is not the preferred rest-frame for timescape cosmology. The line of sight from the observer, who is located in origin of the coordinate system, to one of the groups from the fundamental plane distance catalogue (using the co-moving fundamental plane distance to derive its position in Cartesian coordinates) usually intersects several finite infinity regions. We could have obtained the fraction of the line of sight crossing finite infinity regions semi-analytically. However, this would require calculating all intersections of the lines of sight with the spherical finite infinity regions and applying an interval nesting algorithm afterwards, because the finite infinity regions may overlap. Although this approach would be possible, we found it to be (even after optimisation) very time-consuming and pushing the required computation time far beyond our available resources. Consequently, we had to settle for a slight simplification: we sampled the lines of sight towards each galaxy at 1000 equally spaced locations and counted the fraction of those locations which are within a finite infinity region. We tested this method by comparing the results of the numerical approximation to data calculated semi-analytically for all lines of sight in the observational data. Thereby, we found the additional error of less than 0.2$\%$ produced by the approximation to be sufficiently small compared to all other sources of uncertainties, such as in the reconstruction of the finite infinity regions from observational data. This allowed us to obtain the fraction of the line of sight within finite infinite regions and thereby (along with the previously obtained relative individual Hubble parameters) all the required data that can be derived from our observational data. 

\subsection{Coherent infall and biases}
\label{sec_infall}
\begin{figure}[ht]
\begin{center}
\includegraphics[width=0.45\textwidth]{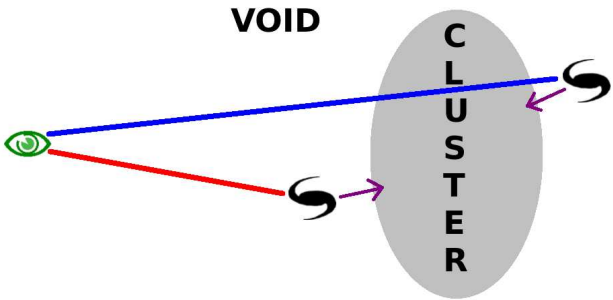}\\ 
\caption{Coherent infall into clusters creating a qualitatively similar signal in redshift space as differential expansion. Galaxies in front of a cluster have fewer overdense regions in the lines of sight than galaxies behind a cluster, simply because then the cluster (which is overdense) is part of their lines of sight.}
\label{infall}
\end{center}
\end{figure}

Importantly for the final analysis, we have to consider which potential effects within $\Lambda$-CDM cosmology may generate a similar effect as the one expected from timescape cosmology. Indeed the coherent infall \citep{Kaiser:1987,Hatton:1998,Coil:2001,Outram:2001} of galaxies into a cluster (see Figure \ref{infall}, which illustrates why one obtains larger redshifts (with respect to their distances) for the galaxies in front of the cluster than for those behind it) naturally creates a signal in redshift space that is qualitatively similar to the effect predicted by timescape cosmology. Thus, one has to estimate the magnitude of this effect very carefully to disentangle it from the signal of timescape cosmology. To this end, we took the mock catalogues, which were used to calibrate and optimize the group finder algorithm in \citet{Saulder:2016}. It was important to pay close attention to the galaxy classification, because the infall effect also depends on it \citep{Coil:2008,Loh:2010} as a consequence of the density-morphology relation \citep{Dressler:1980,Postman:1984,Dressler:1997,Holden:2007,vdWel:2010}. The mock catalogues used in \citet{Saulder:2016} did not consider any galaxy morphology, which was not necessary for their application in that paper. Hence, we had to supplement them with additional data.

Furthermore, various Malmquist biases were affecting our data, because it was derived from a magnitude limited survey with additional limits in redshift and distance. In order to make our comparison between observations and predictions unaffected by Malmquist biases, we implemented in the mock catalogues the same constraints and limits that affect our observational data obtained from SDSS. Thereby, we considered the homogeneous and inhomogeneous Malmquist bias, when comparing the observational data to the simulated one in a later step. It was important to also consider galaxy morphology for these biases, because due to the density-morphology relation \citep{Dressler:1980,Postman:1984,Dressler:1997,Goto:2003,Cappellari:2011} the inhomogeneous Malmquist bias is notable smaller for early-type galaxies than for late-type galaxies.

\subsection{Identifying early-type galaxies in the Millennium simulation}
\begin{figure}[ht]
\begin{center}
\includegraphics[width=0.45\textwidth]{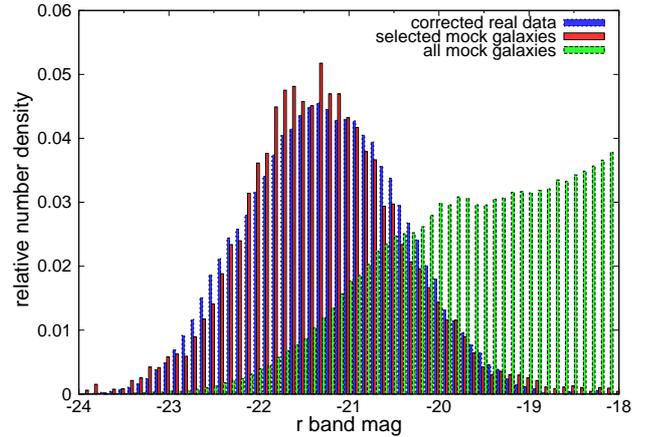}\\ 
\caption{Comparison of the luminosity functions of the observational data and the simulated galaxies used for our mock catalogues. The blue histogram represents the Malmquist-bias corrected luminosity function of the early-type galaxies used to calibrate the fundamental plane. The red histogram shows the selected early-type galaxies from the Millennium simulation with the criteria mentioned in this chapter. The green histogram represents all galaxies from the Millennium simulation for comparison.}
\label{lumfct_combo}
\end{center}
\end{figure}
\begin{figure*}[ht]
\begin{center}
\includegraphics[width=0.30\textwidth]{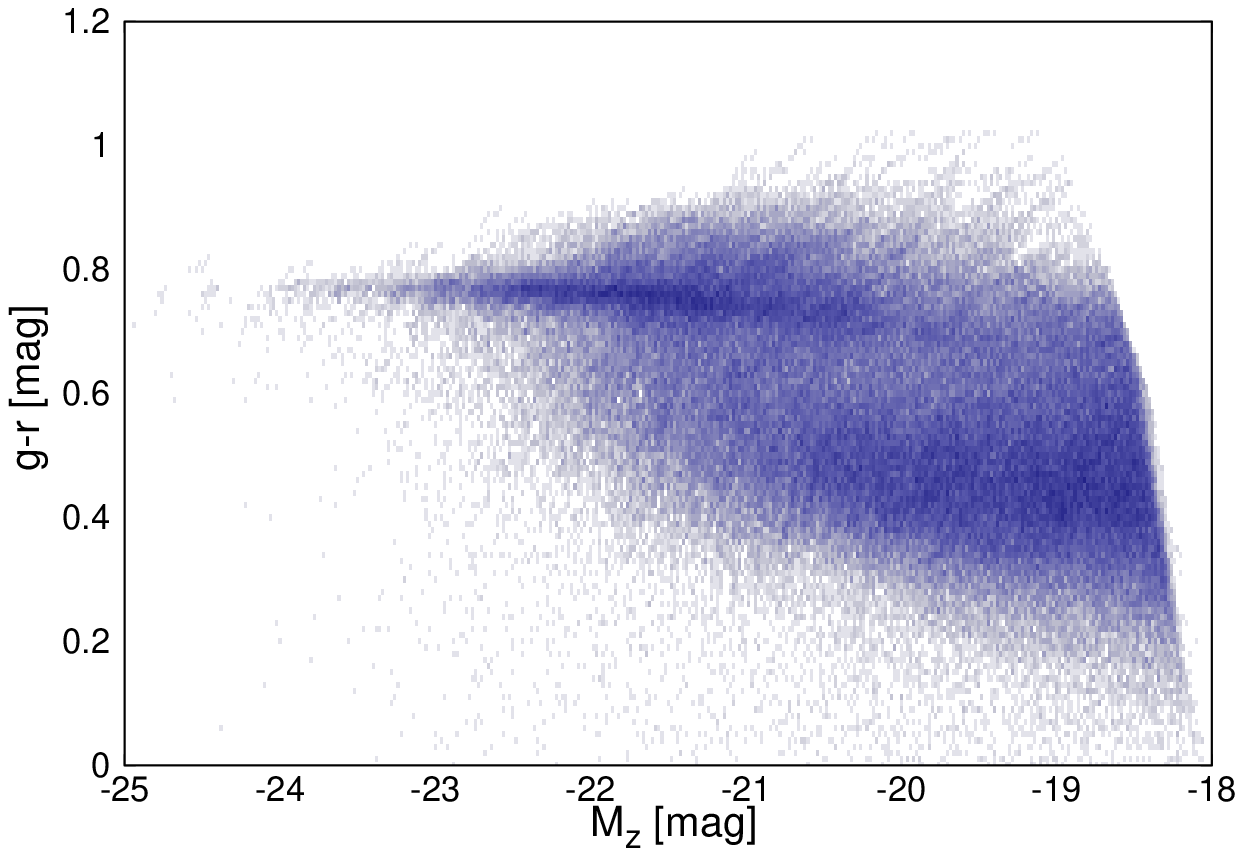}
\includegraphics[width=0.30\textwidth]{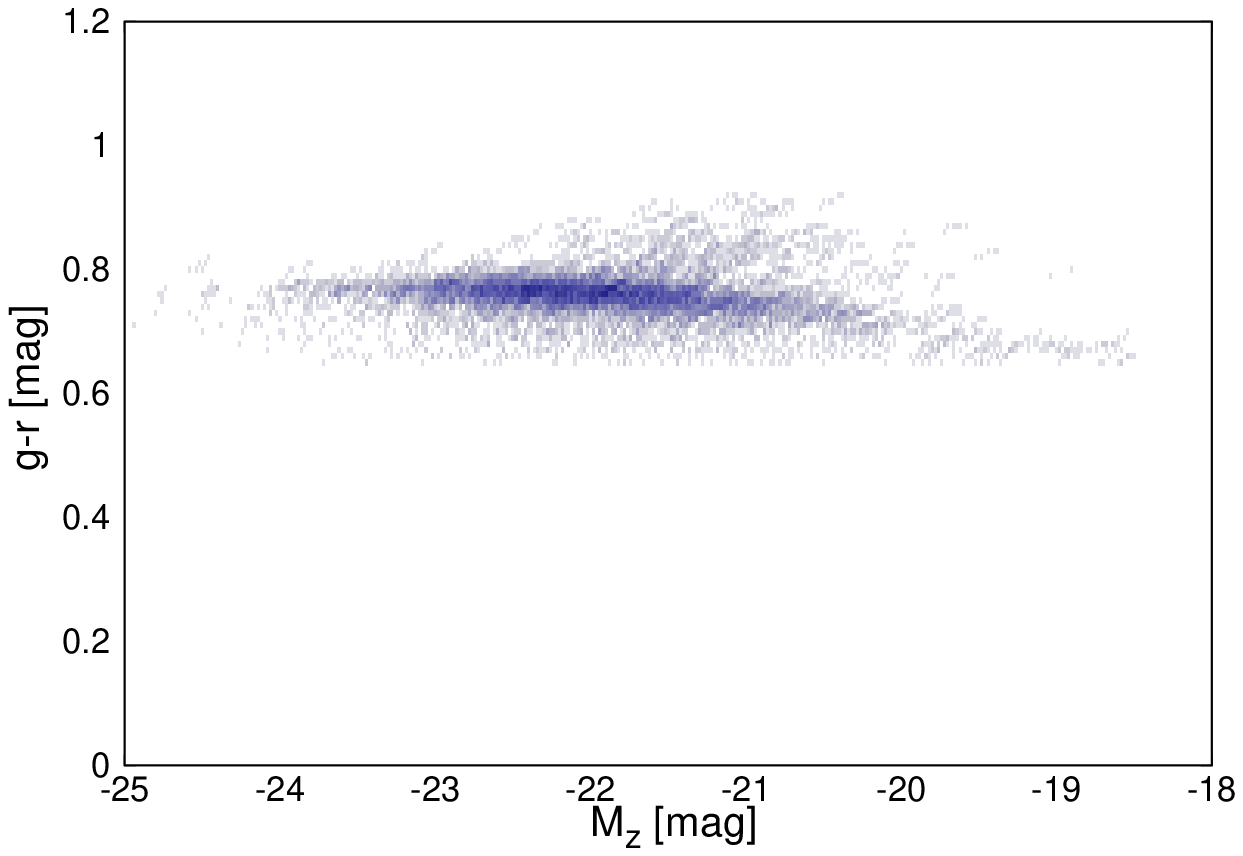}
\includegraphics[width=0.30\textwidth]{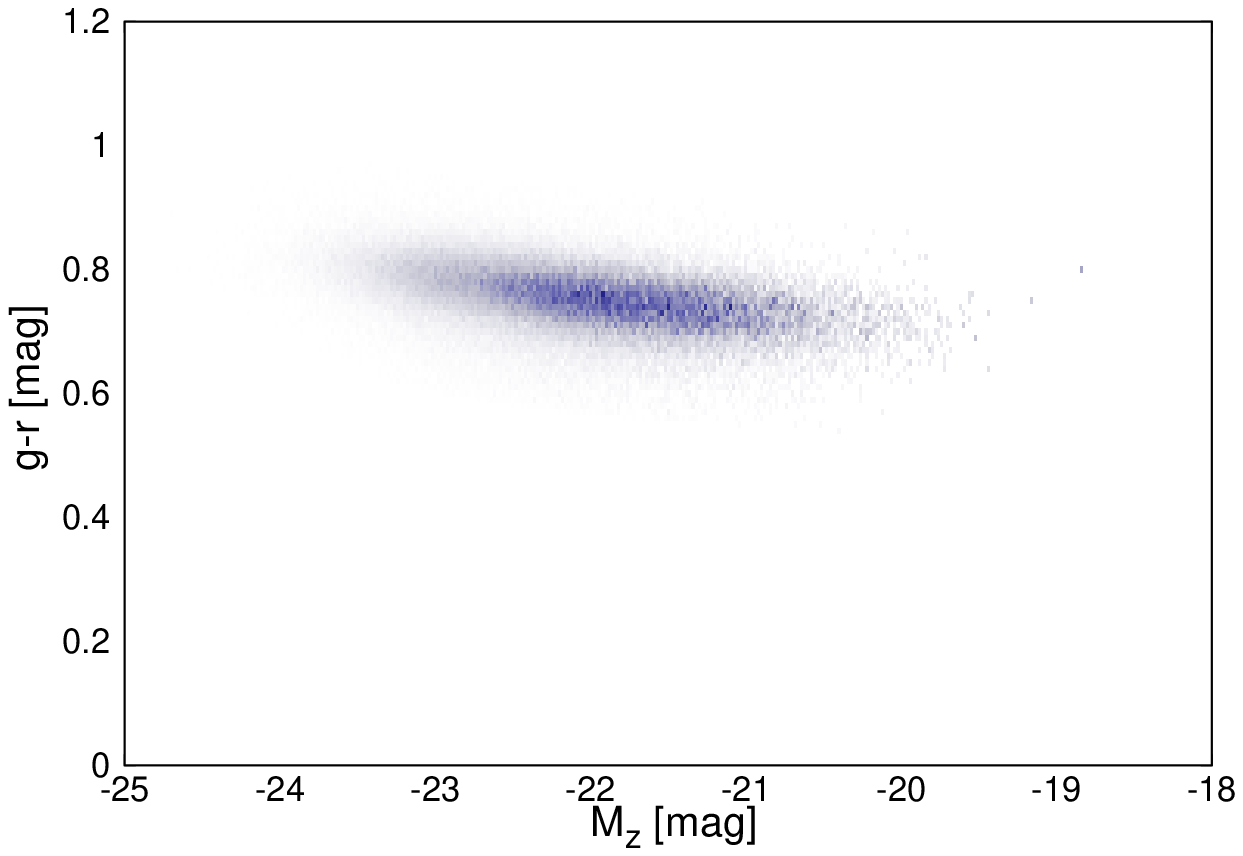}
\caption{Comparison of the colour-magnitude diagrams. First panel: all galaxies with $M_{\textrm{r,SDSS}} < -18$ from the Millennium simulation. Second panel: early-type galaxies selected according to the listed criteria from the Millennium simulation. Third panel: the early-type galaxies from SDSS used to calibrate the fundamental plane.}
\label{CMD_selection}
\end{center}
\end{figure*}

To obtain a suitable mock catalogue for the fundamental plane distances, one has to identify early-type galaxies in the data of the Millennium simulation \citep{millennium,Mill_data}. We thus defined a set of criteria, which yielded a number density of early-type galaxies and distribution in the colour-magnitude diagram similar to our observations. In the observational data \citep{Saulder:2013,Saulder:2015a} we identified early-type galaxies primarily using GalaxyZoo \citep{GalaxyZoo,GalaxyZoo_data}. In the Millennium simulation, dark matter halos are populated by semi-analytical galaxy models \citep{Guo:2010}. These models provide several parameters, which we used for classifications. The following conditions yielded a sample of simulated galaxies, which had a sufficiently similar luminosity function and number density as the observed GalaxyZoo-selected sample of early-type galaxies: A galaxy had to be 
\begin{itemize}
\item bright (SDSS r band magnitude brighter than -18),
\item red\footnote{be part of the red sequence in the colour magnitude diagram} (g-r colour greater than 0.65 mag),
\item poor in cold gas (cold gas to total mass ratio less than 0.0008),
\item quiescent (star formation rate by total mass less than 0.01 per $10^{10}$ years).
\end{itemize}
These values were chosen by an empirical analysis of the data. A comparison of the luminosity functions of the sample of simulated galaxies obtained and the sample of real galaxies used to calibrate the fundamental plane illustrates that this set of criteria yields the desired galaxy sample (see Figure \ref{lumfct_combo}). Furthermore, the colour-magnitude diagram of the hereby selected sample of simulated galaxies shows that red-sequence galaxies are indeed selected and that the distribution of the simulated galaxies is comparable to the distribution of the observed galaxies (see Figure \ref{CMD_selection}). The galaxy density of the simulated galaxy sample is $7.4 \cdot 10^{-4}$ galaxies per Mpc$^{3}$, which is close to value measured for the SDSS sample of $7.9 \cdot 10^{-4}$ galaxies per Mpc$^{3}$. In the end, 733 790 early-type galaxies from the last six snapshots (the same used in \citet{Saulder:2016}) were identified in the Millennium simulation. From this sample, we only considered the galaxies, which were also part of our mock catalogues from \citet{Saulder:2016}. Thereby, it was assured that selection effects and biases of the mock catalogues were adequately considered for our sample of early-type galaxies. 

The only additional effect that had to be taken into account for the early-type galaxies was the scatter in the fundamental plane distance estimate. The fundamental plane distances in the mock catalogues were obtained by using the true distance (the co-moving distance directly from the simulation) and creating a Gaussian scatter around the logarithmized distance values with a width corresponding to the root mean square of the fundamental plane calibration\footnote{It is actually not the root mean square listed in \citet{Saulder:2015a}, but a slightly lower value because of the consideration of the residual redshift dependence.}. The residual redshift dependence of the fundamental plane calibrations was also considered (see \citet{Saulder:2013}). 

\subsection{Creating mock catalogues in $\Lambda$-CDM cosmology}
To get a dataset on which our test can be applied, we ran our group finder algorithm from \citet{Saulder:2016} on the eight mock catalogues of that paper and proceed in the same way as described there to derive the finite infinite regions from the mock catalogues. The previous obtained set of early-type galaxies and their fundamental plane distances was cross-matched with the group catalogues based on the mock catalogues to create a fundamental plane distance group catalogue. By applying the same analysis that we performed for the observational data on these datasets, we received a baseline for the comparison to the prediction of $\Lambda$-CDM cosmology.

\subsection{Creating mock catalogues for timescape cosmology}
The mock observations for timescape cosmology required additional considerations. Ideally, one would want a self-consistent cosmological simulation that considers backreactions from inhomogeneities, but despite recent advances \citep{Racz:2017}, there is currently no numerical simulation with sufficient resolution and volume to create suitable mock-catalogues for our test. Hence, we had to settle for using data from the Millennium simulation and deforming the $\Lambda$-CDM cosmology of said simulation to derive an approximation of the features predicted by timescape cosmology. When doing this, the assumption was used that the matter distribution in the last snapshots (the same ones as used for the $\Lambda$-CDM mock observations) of the Millennium simulation is a reasonably good representation of the large scale matter distribution in the local Universe (variations on small scales do not matter, because of the sizes of the finite infinity regions). The apparent different expansion rates of voids and walls, which is the effect of timescape cosmology that is investigated here, was artificially introduced in the data by systematically adapting the redshifts in the mock catalogues (see next paragraph for details). To correctly consider all the biases, two sets of finite infinity regions were created: one from the mock catalogues as before (actually the same set as used for the $\Lambda$-CDM mock catalogues) and one from the full FoF group catalogue (dark matter halos) directly from the Millennium simulation. For the latter the missing particles\footnote{The FoF groups only host about half the simulation's mass/particles.} were considered in the same way as done in \citet{Saulder:2016} to obtain the total masses of the groups and later the finite infinity regions. The finite infinity region catalogue based on the FoF groups was used to get an as unbiased as possible estimate of the `true' fraction of the line of sight within the finite infinity regions. For the calculation of that value, the unbiased co-moving distances to the groups hosting early-type galaxies were used. In contrast to this, the other mock catalogue contained all the observational biases. However, since our recipe \citep{Saulder:2016} was a reconstruction of finite-infinity regions from observational data instead of the true unbiased matter distribution, it might suffer from an additional systematic error, because it extrapolates beyond the currently studied theoretical implications of timescape cosmology. Since the basis of our approach is a simple two-phase model, we implicitly assume a linear relationship for the dependence of the individual Hubble parameters on the fraction of finite infinity regions in the line of sight. Without a more sophisticated theoretical framework that would give us a clear parametrization of said dependence, our linear model is the prudent choice for the given dataset. Furthermore, any extensions beyond the linear relationship would require a more complex model of the matter distribution in the local Universe than just sizes for finite infinity regions, which would be far beyond the scope of this paper.

We merged the two catalogues after calculating the fraction of the line of sight inside finite regions and the relative individual Hubble parameter (being define as the ratio of the median redshift of the galaxy group by the fundamental plane distance of the galaxy group) for them by using the true fraction of the line of sight within the finite infinity regions to modify the relative individual Hubble parameter of the other mock catalogue in the following empirical approach: first we obtain average fraction of the line-of-slights which are within finite infinity regions $f_{fi}$ from each mock-catalogue: 
\begin{equation}
f_{fi} \cdot H_{0,\textrm{w}} + \left( 1-f_{fi} \right) \cdot H_{0,\textrm{v}} = H_{0,\textrm{av}}.
\label{timescape_approx}
\end{equation}

Since line-of-sight fractions are one-dimensional, the parameter $f_{fi}$ cannot coincide with the three-dimensional wall-volume fraction $f_{w0}$, except in the extreme limit that both parameters are zero or one. Furthermore, the timescape void and wall volume fractions are defined with respect to the volume-average negatively-curved spatial geometry, not with respect to the line-of-sight spherical averages of the dressed geometry. However, it should still be of the same order magnitude, which is with values in the mock catalogues between 0.30 and 0.47 (depending on the mock catalogue) indeed the case. The values derived from reconstructed finite infinity regions based on the mock catalogues are on average by 0.03 greater than the values obtained from the complete dark matter data. The systematic bias is corrected by simply reducing the values from the mock catalogues accordingly. The bare Hubble parameters $H_{0,\textrm{w}}$ according to timescape cosmology is 50.1 km s$^{-1}$ Mpc$^{-1}$ \citep{Duley:2013}, which sets an approximate value of the expansion rate inside finite infinity regions. The dressed Hubble parameter $H_{0,\textrm{av}}$ in timescape cosmology is 61.7 km s$^{-1}$ Mpc$^{-1}$, which corresponds to the value measured by an observer inside a galaxy (such as ourselves). Within our approximated timescape model, it would correspond to the average measured Hubble parameter. Hence, we rescale the values of the mock catalogues, which in their creation assumed a $\Lambda$-CDM cosmology with a (universal) Hubble parameter of 73 km s$^{-1}$ Mpc$^{-1}$ \citep{millennium}, to the dressed value. Using equation \ref{timescape_approx}, we can derive the expansion rate of the voids $H_{0,\textrm{v}}$ for our deformed $\Lambda$-CDM model, which we use to approximate timescape cosmology (approximated timescape model, in short). With the help of all these parameters, we modify individual Hubble parameters from the mock catalogues according to the fraction of finite infinity regions (based on the dark matter halo catalogue) in the line of sight. For example, the individual Hubble parameter is reduced by 11.6 s$^{-1}$ Mpc$^{-1}$ in the case of a galaxy group observed entirely through finite infinity regions or increased to much higher values for galaxy groups with primarily void environment in their lines of sight. When writing the code for this calibration, we also added an additional multiplicative parameter $b_{\textrm{soft}}$, which allowed us to adjust the introduced slope to study a greater range of backreaction scenarios. This parameter is set to 0.75 to account for the systematic and statistical biases as well as the error bars of the predicated different expansion rates. A $b_{\textrm{soft}}$ of zero would remove the differential expansion completely, while a $b_{\textrm{soft}}$ of one would keep it the values derived from the best fit values of \citet{Duley:2013} and our estimated average fraction of the lines of sight which are within finite infinity regions for each specific mock catalogue used. In the last step, we divide all the individual Hubble parameters by the average Hubble parameter to allow easy comparison between the datasets of different cosmologies. These renormalized relative individual Hubble parameters together with the fractions of the line of sight inside finite infinity regions (the values from the mock catalogue, not the `true' values from dark matter halos) constituted the dataset, which represents the closest approximation of timescape cosmology that we can achieve with the available data. We are fully aware that this model is far from perfect, but it should provide sufficient information to indicate the direction of deviations due to timescape cosmology from the $\Lambda$-CDM cosmology in the observational data as well as a rough estimate of their magnitude.

With the three different datasets (the observational data, the $\Lambda$-CDM mock data, and the differential expansion mock data, which we used to approximated timescape cosmology) at hand, we could execute our test.

\section{Results}
\label{sec_res}

\subsection{Mock catalogues}
\begin{figure}[ht]
\begin{center}
\includegraphics[width=0.45\textwidth]{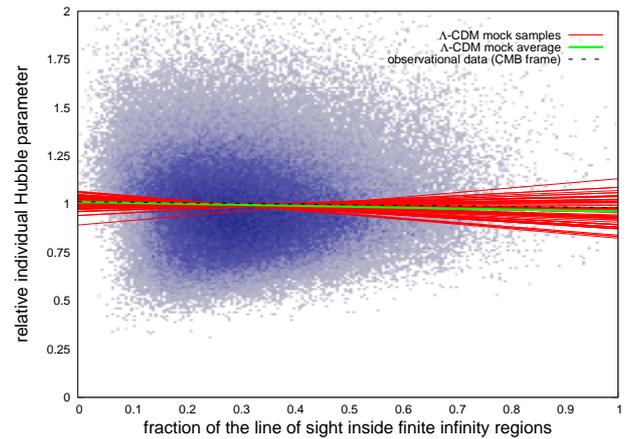}\\ 
\caption{Diversity of the combined $\Lambda$-CDM mock catalogues. Red lines: weighted fits on combined mock catalogues; green line: the average obtained from theses fits; bluish point cloud: distribution of all galaxy groups hosting early-type galaxies weighted by number of early-type galaxies in them from all mock catalogues.}
\label{result_scatter_lcdm}
\end{center}
\end{figure}

The content of the eight mock catalogues using $\Lambda$-CDM cosmology form the baseline for the cosmological test presented here. Any significant deviations of the observational data from these results would challenge our current understanding of the Universe. The area covered by one mock catalogue is by design one eighth of the entire sky, while the real observations cover almost one quarter of the sky ($\sim$22.7$\%$ \citep{SDSS_DR12}). To improve the comparability between the mock catalogues and the observations, two mock catalogues of the same cosmology were combined in the following analysis. The combined mock catalogues cover $25\%$ of the sky, which is reasonably close to the SDSS spectroscopic sky coverage and allowed for a direct comparison between the combined mock catalogues and the observations. Hence, a sample of 64\footnote{Considering all possible double combinations of the 8 mock catalogues.} combined mock catalogues was created of which 36 were unique combinations (28 are combinations of two different mock catalogues, of which each appeared twice in the sample, while the eight combinations of the same mock catalogues with itself only appeared once, thereby providing the correct statistical weights for the likelihood of the combinations). Because each combined mock catalogue covered an area of the sky comparable to the observations, we were able to derive probabilities and statistics for expectations of $\Lambda$-CDM cosmology. In Figure \ref{result_scatter_lcdm}, the weighted (by the square root of the number of group members) linear regression fits on the distributions of relative individual Hubble parameters and fractions of the line of sight within finite infinity regions obtained from the combined mock catalogues are displayed. We also calculated the average expectation for the slope using the mean values of the coefficients obtained by the fits on the different combined mock catalogues. 

\begin{figure}[ht]
\begin{center}
\includegraphics[width=0.45\textwidth]{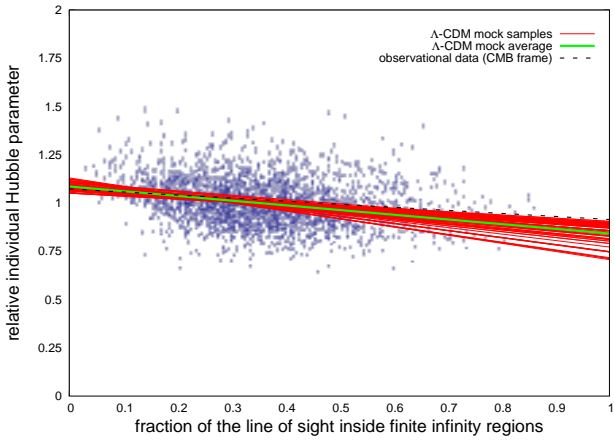}\\ 
\caption{Diversity of the combined $\Lambda$-CDM mock catalogues. Red lines: weighted fits on a selected sub-sample (at least 3 early-type galaxies and a fundamental plane distance of less than 402.8 Mpc) of the combined mock catalogues; green line: the average obtained from theses fits; bluish point cloud: distribution of all galaxy groups hosting early-type galaxies weighted by number of early-type galaxies in them from a selected sub-sample of all mock catalogues.}
\label{result_scatter_lcdm_lim}
\end{center}
\end{figure}

To reduce the scatter, which is primarily caused by the intrinsic scatter in the fundamental plane distance measurements to individual galaxies, a sub-sample of galaxy groups was selected using the following criteria: at least three early-type galaxies in the group\footnote{The mean values of the parameters (redshift, fundamental plane distance, ...) of the group were used.}, and a co-moving fundamental plane distance of less than 402.8 Mpc (which corresponds to the redshift limit of 0.1 for the cosmology of the Millennium simulation). As illustrated in Figure \ref{result_scatter_lcdm_lim}, these criteria visibly reduced the scatter of the weighted least-square fits. Removing the wide scatter due to the relatively high uncertainties in the fundamental plane distances of galaxy groups with only one or two early-type galaxies also caused the fitted slope to be steeper than for the full data. The disadvantage of this procedure is that it reduces the size of sample. This is not unproblematic, because there is a dearth of rich groups in all mock catalogues based on the Millennium simulation compared to observed data. Hence, any criterion selecting groups based on their richness reduced the sample size of the mock catalogues disproportionally compared to SDSS data. This issue with the mock catalogues is inherent in the Millennium simulation and grows worse when looking for groups hosting early-type galaxies. There is little that can be done to avoid it and it had to be considered as a systematic bias. 

\begin{figure}[ht]
\begin{center}
\includegraphics[width=0.45\textwidth]{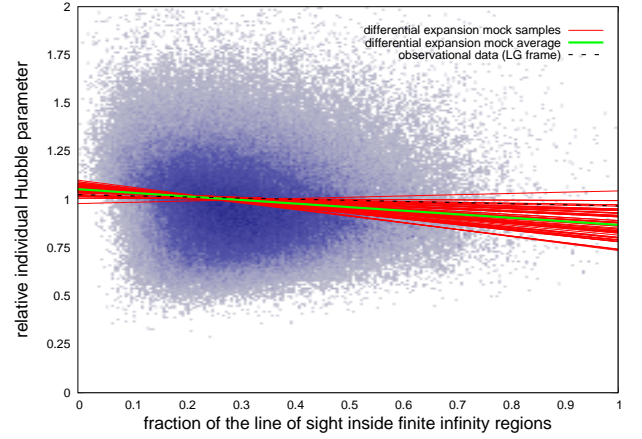}\\ 
\caption{Diversity of the combined differential expansion mock catalogues. Red lines: weighted fits on combined mock catalogues; green line: the average obtained from theses fits; bluish point cloud: distribution of all galaxy groups hosting early-type galaxies weighted by number of early-type galaxies in them from all mock catalogues.}
\label{result_scatter_ts}
\end{center}
\end{figure}
\begin{figure}[ht]
\begin{center}
\includegraphics[width=0.45\textwidth]{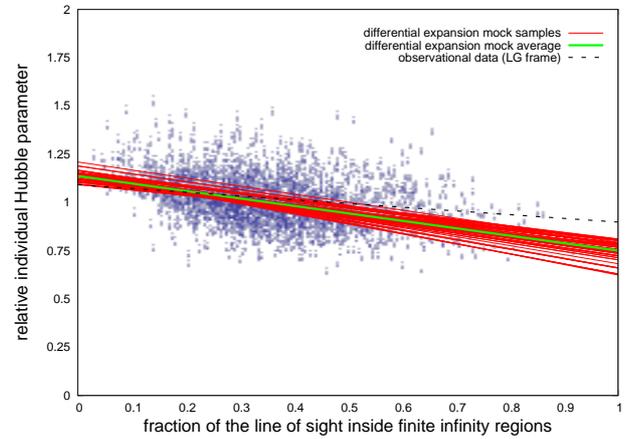}\\ 
\caption{Diversity of the combined differential expansion mock catalogues. Red lines: weighted fits on a selected sub-sample (at least 3 early-type galaxies and a fundamental plane distance of less than 402.8 Mpc) of the combined mock catalogues; green line: the average obtained from theses fits; bluish point cloud: distribution of all galaxy groups hosting early-type galaxies weighted by number of early-type galaxies in them from a selected sub-sample of all mock catalogues.}
\label{result_scatter_ts_lim}
\end{center}
\end{figure}

We repeated the same analysis as done for $\Lambda$-CDM mock catalogues with the differential expansion mock catalogues. The overall distributions (see Figures \ref{result_scatter_ts} and \ref{result_scatter_ts_lim}) are similar to the previous ones, but the fits show a clearly steeper slope for the individual Hubble parameters depending on the fraction of the line of sight within finite infinity regions (wall environment). This is due to the additional effect of the different expansion rates of voids and walls in timescape cosmology, which we considered in our model and aimed to find in the observational data when executing our test. 

\subsection{Linear regression}
\begin{figure}[ht]
\begin{center}
\includegraphics[width=0.45\textwidth]{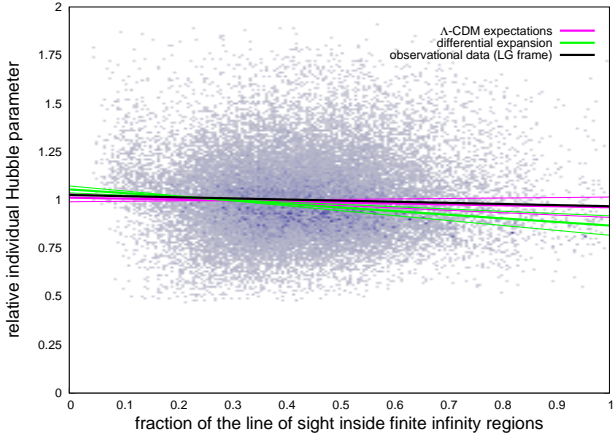}\\ 
\caption{Full observational data compared to the model predictions in the Local Group (LG) rest frame. Black line: weighted least-square fit on the observed data in the LG rest frame; green lines: expectations and 3-$\sigma$ limits of the $\Lambda$-CDM model; magenta lines: expectations and 3-$\sigma$ limits of the differential expansion model.}
\label{map_obs_lg}
\end{center}
\end{figure}

\begin{figure}[ht]
\begin{center}
\includegraphics[width=0.45\textwidth]{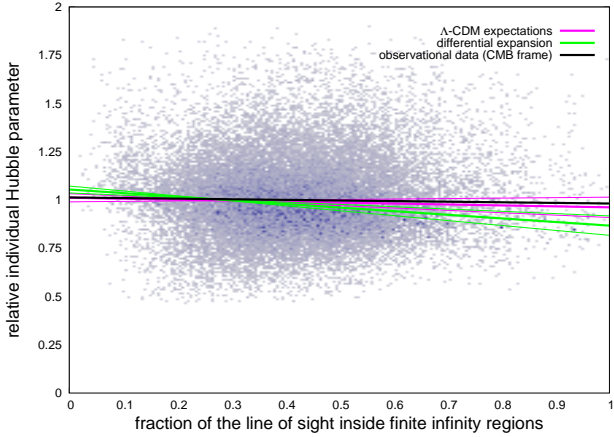}\\ 
\caption{Full observational data compared to the model predictions in the Cosmic Microwave Background (CMB) rest frame. Black line: weighted least-square fit on the observed data in the CMB rest frame; the other symbols are the same as in Figure \ref{map_obs_lg}.}
\label{map_obs_cmb}
\end{center}
\end{figure}

\begin{figure}[ht]
\begin{center}
\includegraphics[width=0.45\textwidth]{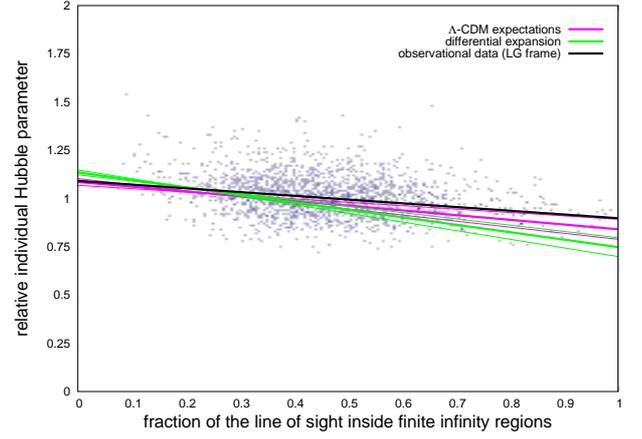}\\ 
\caption{Key Results Ia: selected groups (at least 3 early-type galaxies and a fundamental plane distance of less than 402.8 Mpc) compared to the model predictions in the Local Group (LG) rest frame. Black line: weighted fits on a selected sub-sample of the observed data in the LG rest frame; the other symbols are the same as in Figure \ref{map_obs_lg}.}
\label{map_obs_lg_selected}
\end{center}
\end{figure}

\begin{figure}[ht]
\begin{center}
\includegraphics[width=0.45\textwidth]{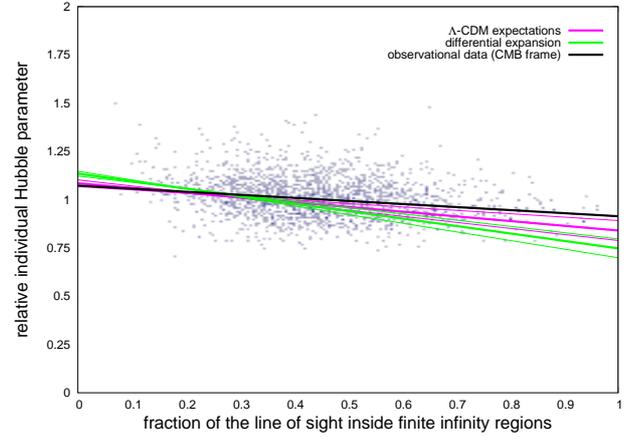}\\ 
\caption{Key Results Ib: selected groups (at least 3 early-type galaxies and a fundamental plane distance of less than 402.8 Mpc) compared to the model predictions in the Cosmic Microwave Background (CMB) rest frame. Black line: weighted fits on a selected sub-sample of the observed data in the CMB rest frame; the other symbols are the same as in Figure \ref{map_obs_lg}.}
\label{map_obs_cmb_selected}
\end{center}
\end{figure}

\begin{table*}
\begin{center}
\begin{tabular}{ccccccc}
model & frame &$k_{\textrm{obs}}$ & $\overline{k_{\textrm{mock}}}$ & $\sigma_{k,\textrm{mock}}$ & $\Delta k [\sigma]$ & $P$ \\ \hline
 $\Lambda$-CDM (full)     & LG  &-0.06 & -0.05 & 0.25 & 0.34  & 0.734  \\
 $\Lambda$-CDM (full)     & CMB &-0.03 & -0.05 & 0.25 & 0.74  & 0.459  \\
 $\Lambda$-CDM (selected) & LG  &-0.19 & -0.25 & 0.23 & 2.24  & 0.025  \\ 
 $\Lambda$-CDM (selected) & CMB &-0.16 & -0.25 & 0.23 & 3.75  & 0.0002  \\
 differential expansion (full)    & LG  &-0.06 & -0.19 & 0.23 & 5.69  & $< 0.0001$ \\
 differential expansion (full)    & CMB &-0.03 & -0.19 & 0.23 & 6.86  & $< 0.0001$ \\
 differential expansion (selected)& LG  &-0.19 & -0.39 & 0.20 & 9.50 & $< 0.0001$ \\
 differential expansion (selected)& CMB &-0.16 & -0.39 & 0.20 & 11.22 & $< 0.0001$
\end{tabular}
\end{center}
\caption{Coefficients, probabilities and likelihoods of the observations fitting the different models and rest frames (LG stands for local group and CMB for cosmic microwave background). $k_{\textrm{obs}}$: slope of the weight least square fitted linear regression on the observational data; $\overline{k_{\textrm{mock}}}$: mean expected slope for a specific model; $\sigma_{k,\textrm{mock}}$: standard deviation of the slope for a specific model (derived using the different combined mock catalogues); $\Delta k$: deviation of the observations from the mean value of the model, given in standard deviations; $P$: probability for the observations to be represented by the model.}
\label{results_k}
\end{table*}

The next step of our analysis was to compare the observations in two different rest frames to the two mock datasets. The Comic Microwave Background (CMB) rest frame is the natural choice within $\Lambda$-CDM cosmology, while the Local Group (LG) rest frame is an appropriate choice within the framework of timescape cosmology being close to our own finite infinity region.

We identified nearly 30 000\footnote{29 527 for the CMB rest frame and 29 717 for the Local Group rest frame. The small difference in the numbers is due to the redshift cut, which close to its upper limit does not cover exactly the same volume in both frames.} groups hosting at least one early-type galaxies within a redshift range of 0 and 0.1 distributed over 9376 square degree \citep{SDSS_DR12} in SDSS DR12. In the same area, but within a redshift range of 0 and 0.11, we identified almost 160 000\footnote{158 607 for the CMB rest frame and 157 580 for the LG rest frame.} finite infinity regions. Each of the combined mock catalogues contained between 35 514 and 41 238 groups hosting early-type galaxies and between 190 176 and 233 170 finite infinity regions for the $\Lambda$-CDM data and between 9 215 348 and 9 217 362 finite infinity regions based on the true dark matter data, which we used in constructing our differential expansion model distributed over about 10 313 square degree\footnote{One quarter of the entire sky.} within the same redshift limits as for the observed data. The groups hosting early-type galaxies, with their relative individual Hubble parameters (derived from the group's median redshift and fundamental plane distance) and fractions of the finite infinity regions in the line of sight, form the data points in the subsequent analysis. As illustrated in Figures \ref{map_obs_lg} and \ref{map_obs_cmb} and shown in Table \ref{results_k}, the slope of the weighted least square fitted linear\footnote{A linear fit seems well-suited, because our mock catalogues for the differential expansion model assumed a linear gradient.} regression on the observational data is within 1-$\sigma$ of the expectations of $\Lambda$-CDM cosmology, while the differential expansion model is off by more than 2-$\sigma$ even its preferred LG rest frame. The absolute zero point of the linear regression is not essential for the analysis, because it is extremely sensitive to the normalization of the data and the exact value of $f_{fi}$ and does not allow for a proper distinction between the two cosmological theories. 

To reduce the relatively large scatter (mostly originating from the uncertainties in the fundamental plane distance estimate) from the different mock catalogues, we decided to use the selected sample for the rest of our analysis. The size of our sample was diminished to slight more than 1800\footnote{1835 for the CMB rest frame and 1844 for the LG rest frame.} galaxies groups for the observational data and to between 530 and 954 galaxy groups for the combined mock catalogues (the number of finite infinity regions in the models remained unchanged). Due to the already mentioned dearth of rich clusters in the Millennium simulation, the values of mock catalogues were significantly lower than one would expect based on numbers from the observational data. Despite disproportionally reducing the sample size, the uncertainty in the expected slopes was reduced, which allowed us to continue the analysis.

Figures \ref{map_obs_lg_selected} and \ref{map_obs_cmb_selected} thus show a key result of our test: the slope of the fit on the observational data is even flatter than the average expectations of $\Lambda$-CDM cosmology. At the same time, this results is a massive outlier (see Table \ref{results_k}) for the expectations of the differential expansion model, showing that the preference for $\Lambda$-CDM cosmology is even stronger when using the selected sample.

\subsection{Analysis of binned data}
\begin{figure}[ht]
\begin{center}
\includegraphics[width=0.45\textwidth]{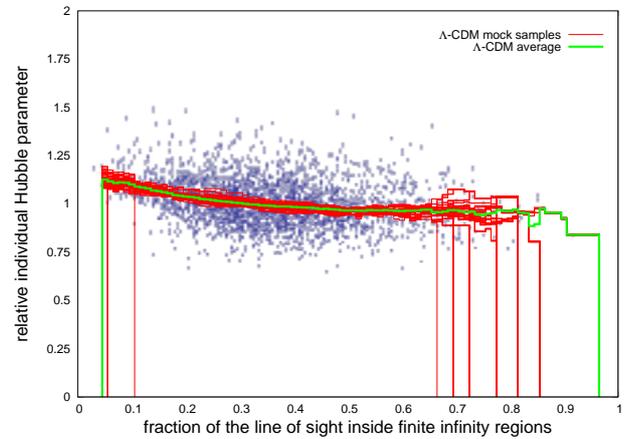}\\ 
\caption{Diversity of the combined $\Lambda$-CDM mock catalogues. Red lines: weighted average in the bins of each combined mock catalogues only containing a selected sub-sample of groups; green line: the average in the bins of all combined mock catalogues; bluish point cloud: distribution of a selected sub-sample of groups in all mock catalogues.}
\label{result_bin_lcdm_lim}
\end{center}
\end{figure}

\begin{figure}[ht]
\begin{center}
\includegraphics[width=0.45\textwidth]{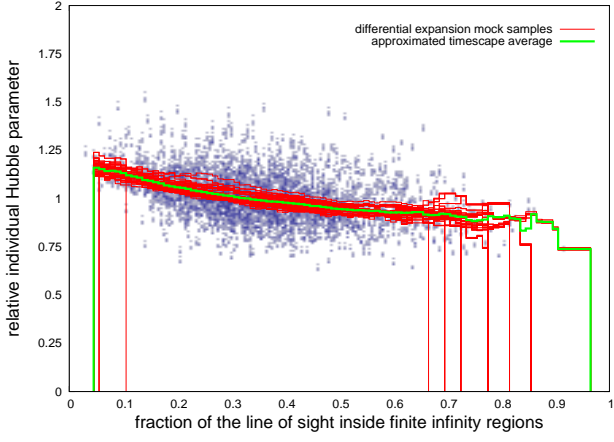}\\ 
\caption{Diversity of the combined differential expansion mock catalogues. Symbols are the same as in Figure \ref{result_bin_lcdm_lim}}
\label{result_bin_ts_lim}
\end{center}
\end{figure}

Aside from a least square fitted linear regression, we performed an analysis on the data in which we split the data into bins corresponding to different ranges in the fraction of the line of sight within finite infinity regions. Each bin was 0.1 wide and we shifted their centre iteratively by 0.01 from 0.05 to 0.95 along the axis of the fraction of the line of sight within finite infinity regions. Within these bins the weighted averages of the relative individual Hubble parameters were calculated for each combined mock catalogue. In the case of the selected sample, the outermost bins were empty for some of the combined mock catalogues (see Figures \ref{result_bin_lcdm_lim} and \ref{result_bin_ts_lim}) due to already mentioned dearth of rich groups in the mock catalogues. The scatter between the different combined mock catalogues increases at both extremes due to the poorer sampling there. 

\begin{figure}[ht]
\begin{center}
\includegraphics[width=0.45\textwidth]{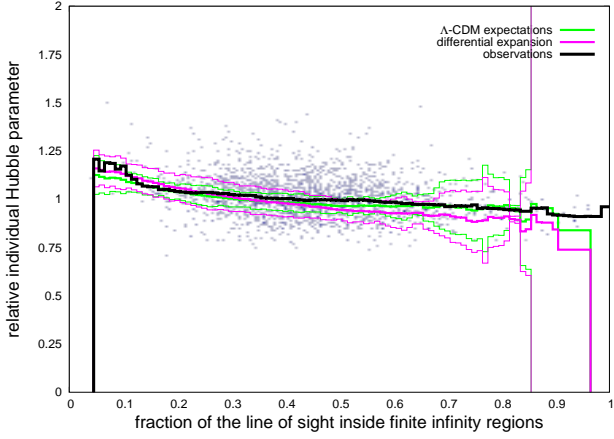}\\ 
\caption{Selected observational data in the CMB rest frame compared to the model predictions. Black line: weighted average in the bins of the observed data; green lines: expectations and 3-$\sigma$ limits of the $\Lambda$-CDM model for the bins; magenta lines: expectations and 3-$\sigma$ limits of the differential expansion model for the bins.}
\label{finalresults_bins_lim_cmb}
\end{center}
\end{figure}

\begin{figure}[ht]
\begin{center}
\includegraphics[width=0.45\textwidth]{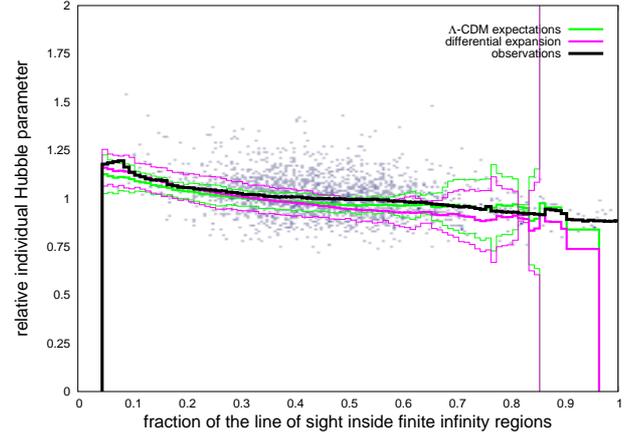}\\ 
\caption{Selected observational data in the LG rest frame compared to the model predictions. Symbols are the same as in Figure \ref{finalresults_bins_lim_cmb}}
\label{finalresults_bins_lim_lg}
\end{center}
\end{figure}

We compared the values in the bins predicted by the two models to the observations in the CMB and LG rest frame (see Figures \ref{finalresults_bins_lim_cmb} and \ref{finalresults_bins_lim_lg}). In the CMB frame 47 of 90 bins have a higher probability for the $\Lambda$-CDM model, while in the LG frame 57 of 90 bins show a higher probability for the differential expansion model. However given the large scatter (the observational data lies within the error bar of both theories) and the sensitivity to scaling, there is little significance in these results. Furthermore, the observed data lies well within the error bars of both models in all cases.

\begin{figure}[ht]
\begin{center}
\includegraphics[width=0.45\textwidth]{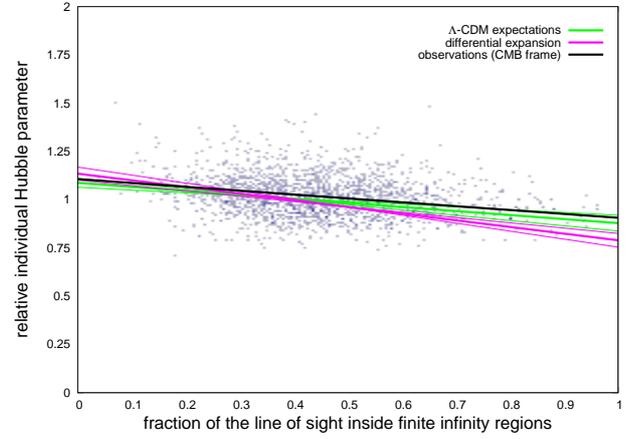}\\ 
\caption{Key Result IIa, using binned analysis: selected groups in the CMB rest frame compared to the model predictions. Black line: linear regression fit on the equally weighted bins of the observed data; green lines: averaged linear regression fit on the binned data of the $\Lambda$-CDM mock catalogues; magenta lines:averaged linear regression fit on the binned data of the timescape mock catalogues. The bins are the same as in Figures \ref{result_bin_lcdm_lim} and \ref{result_bin_ts_lim}, but are not explicitly shown to keep the plot clean.}
\label{binfit_selected_cmb}
\end{center}
\end{figure}

\begin{figure}[ht]
\begin{center}
\includegraphics[width=0.45\textwidth]{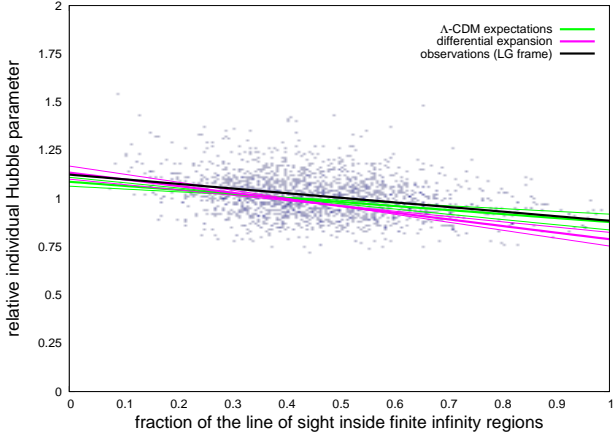}\\ 
\caption{Key Result IIb, using binned analysis: selected groups in the LG rest frame compared to the model predictions. Symbols are the same as in Figure \ref{binfit_selected_cmb}}
\label{binfit_selected_lg}
\end{center}
\end{figure}

\begin{table*}
\begin{center}
\begin{tabular}{cccccccc}
model & frame & $k_{\textrm{obs}}$ & $k_{\textrm{mock}}$ & $\sigma_{k,\textrm{mock}}$ & $\Delta k [\sigma]$ & $P$ & $\mathcal{L}$ \\ \hline
$\Lambda$-CDM  & LG  & -0.24 & -0.21 & 0.06 & 0.48 & 0.63 & 0.83 \\
$\Lambda$-CDM  & CMB & -0.20 & -0.21 & 0.06 & 0.12 & 0.90 & 0.96 \\
differential expansion & LG  & -0.24 & -0.34 & 0.07 & 1.52 & 0.13 & 0.17 \\
differential expansion & CMB & -0.20 & -0.34 & 0.07 & 2.08 & 0.04 & 0.04
\end{tabular}
\end{center}
\caption{Linear regression to the binned data (see Figures \ref{binfit_selected_cmb} and \ref{binfit_selected_lg}). Coefficients, probabilities and likelihoods of the observations using the selected datasets calculated for the average of the linear regressions fitted to the bins. First column: specific model; second column: rest frame; third column: slope $k_{\textrm{obs}}$ using the observations; fourth column: slope $H_{\textrm{mock}}$ fitted on the mock catalogue data; fifth column: standard deviation $\sigma_{k,\textrm{mock}}$ of the slope fitted to the mock catalogue ; sixth column: deviation of the observed slope from the slope in the mock catalogue (given in standard deviation); seventh column: probability $P$ for the observations to be represented by the model; eighth column: likelihood $\mathcal{L}$ for the model (only assuming $\Lambda$-CDM and timescape cosmology as possible options).}
 \label{results_binfitlin}
\end{table*}

To get stronger constraints from our analysis of the binned data, we used the following set of data points for each combined mock catalogue: the central value for the ratio of the finite infinity regions in the line of sight of each bin and the average individual Hubble parameter within those bins. For every combined mock catalogue, we fitted a linear regressions on the available data points (every data point was assigned the same statistical weight). For both cosmologies and both rest frames (CMB and LG), we calculated the average values of the slopes obtained by these fits and the corresponding standard deviation and compared it to the observational data, which was treated in the same way. As illustrated in Figures \ref{binfit_selected_cmb} and \ref{binfit_selected_lg}, the observations follow the expectation of the $\Lambda$-CDM model relatively closely. Due to the use of a different fitting method that avoids putting much weight on the centre of distribution, but considers each binned region with the same significance, there are some differences in the gradient and the scatter of the fit (when compared to Figures \ref{map_obs_lg_selected} and \ref{map_obs_cmb_selected}) for the predictions based on simulations as well as the observational data. The detailed data in Table \ref{results_binfitlin} shows a clear preference for $\Lambda$-CDM cosmology over our approximation of timescape cosmology. The results are not as striking as with the direct linear regression in the previous section, leaving a slim chance for an even more strongly softened\footnote{With lower values of $b_{\textrm{soft}}$.} differential expansion model with a notably flatter slope, especially in the Local Group rest frame. 

\subsection{Kolmogorov-Smirnov test}
As a complementary sanity check, we also performed a Kolmogorov-Smirnov test (KS test) on our data. We used the implementation from the Numerical Recipes \citep{Numerical_recipes} for the two-sample case of the two-dimensional KS test. We compared the observed data to each pair\footnote{A $\Lambda$-CDM mock catalogue and a differential expansion mock catalogue derived from the same combination of mock-catalogues, hence covering the same structures} of $\Lambda$-CDM and differential expansion mock catalogues. For most pairs, we obtained very low probabilities (over a wide range of magnitudes down as low as $10^{-56}$) that the observations are not in conflict with anyone of the two models. The highest probabilities/significance levels produced by any pair are 0.002 for $\Lambda$-CDM cosmology and 0.0002 for the differential expansion model. For all of the 36 different sets of mock catalogues $\Lambda$-CDM cosmology has a higher probability (between a factor of 2 and 27 000 000) than differential expansion mock catalogues according to the KS test. Overall the KS test is consistent with the finding from the previous sections that $\Lambda$-CDM cosmology provides a better fit to the data than our approximation of timescape cosmology. However, the very low significance values made us wonder if that test is ill-suited for our analysis or the implementation, based on the Numerical recipes \citep{Numerical_recipes}, is faulty (although with simple test examples, the code yields reasonable values).

\subsection{$\mathcal{X}^{2}$ test}
We also performed $\mathcal{X}^{2}$ tests by comparing the observational data to each mock-catalogue individually. For the full dataset, the $\mathcal{X}^{2}$ of the differential expansion model is always between 0.1 and 13$\%$ (typically around 1$\%$) higher than that of the corresponding $\Lambda$-CDM model for both rest frames of the observational data. For the selected dataset 13 in the case of the CMB rest frame and 20 in the case of the LG rest frame mock catalogues yield lower $\mathcal{X}^{2}$ values (by less than 4$\%$) for the differential expansion model, while the remainder of the 64 mock catalogues prefers the $\Lambda$-CDM model by up to 32$\%$ (for certain mock catalogues). Although overall the results of $\mathcal{X}^{2}$ test shows a preference for $\Lambda$-CDM cosmology over our approximation of timescape cosmology, the significance is still very low and certain mock-catalogues are still compatible with the observations.

\section{Discussion}
\subsection{Comparison to preliminary data}
In general, the results of the cosmological test performed in this paper showed a preference for $\Lambda$-CDM cosmology over our differential expansion model, which we used to approximate timescape cosmology. This contradicts our preliminary results presented in \citet{Saulder:2012}. Hence, it is important to have a closer look on the changes made between the initial design of the test and its final execution in this paper to better understand the difference. 

The model of the matter distribution of the local Universe in \citet{Saulder:2012} was still very basic: a fixed mass-to-light ratio was assumed for every SDSS galaxy instead of a proper mass estimate covering the entire range from individual galaxies to galaxy clusters as done in \citet{Saulder:2016}. However in unpublished intermediate results\footnote{These results were only presented as a poster at the Ripples in the Cosmos conference in Durham, July 2013: \url{http://www.equinoxomega.net/files/talks_posters/poster_Durham.pdf}.}, an already improved model of the matter distribution \citep{Yang:2007,Yang:2009} in the local Universe was used and we found a slope, which was largely in agreement with the timescape model. The most significant disadvantage of the data then was that it did not contain any galaxies below a redshift of 0.01 and these nearby regions are important for a solid test of timescape cosmology versus $\Lambda$-CDM cosmology. 

This was a main motivation to produce the data presented in \citet{Saulder:2016}, in which we used 2MRS data to complete\footnote{The SDSS spectroscopic sample suffers from a saturation bias, which excludes bright nearby galaxies.} an SDSS based model of the matter distribution in the local Universe. Furthermore, the total amount of matter in the \citet{Yang:2009} catalogue was only a fraction of which would be expected for that volume based on cosmological observations, even at the lower redshifts. At that time, rescaling the masses of that catalogue to the expected value did not take into account the distribution of the matter correctly. In addition to that, the same issues were present with the mock catalogues based on the Millennium simulation for the intermediate data. In \citet{Saulder:2016}, we corrected those issues and provided a significantly better model of the matter distribution in the local Universe as well as better calibrated mock catalogues. In this paper, we provide (see Appendix \ref{newfiregions}) a slightly improved version of that model using more recent values \citep{Duley:2013} for the critical density in timescape cosmology to construct the finite infinity regions.

In \citet{Saulder:2012}, we naively assumed that there would be no slope in the data for $\Lambda$-CDM cosmology, because we were not yet taking into account coherent infall (see Figure \ref{infall}) and other biases from the peculiar motions. Nevertheless, the observed data showed a much steeper slope of the linear regression fitted to the data than even the expectation of our simplified timescape cosmology-based model (which was not taking into account that bias either). Even considering the effect of peculiar motions, we still found a slope steeper than what would be expected for our approximation of timescape cosmology. In the final data presented here, all tools and catalogues used in test were calibrated consistently and we used a complete model of the matter distribution in the local Universe. The steepness of the slope of the individual Hubble parameters depending on the fraction of the line of sight within finite infinity regions is affected by normalization as well as the total amount of matter in the calibrations, because it affects the sizes of the finite infinity regions and consequently the total volume covered by them. Hence, our empirical approach of finding the average fraction of the line of sight within finite infinity regions (see Equation \ref{timescape_approx}) would be directly affected by notable changes in their sizes. The self-consistency of the calibrations was definitely the most important improvement from our first results in \citet{Saulder:2012} to our final results presented here. The fundamental plane calibrations were another improvement since our preliminary results. The fundamental plane calibrations used in \citet{Saulder:2012} did not consider the Malmquist-bias and other effects correctly. This was fixed in \citet{Saulder:2013}, in which the fundamental plane was properly calibrated using a huge SDSS sample. These calibrations were already used for the intermediate data, but did not significantly alter the overall outcome that much. An extension of the sample size and further improvements on the fundamental plane calibration were made in \citet{Saulder:2015a}, which we used to obtain the final results. Considering all the changes and improvement, the differences between the first results \citep{Saulder:2012} and the results presented here appear comprehensible.

\subsection{Approximations for timescape cosmology}
\begin{figure}[ht]
\begin{center}
\includegraphics[width=0.45\textwidth]{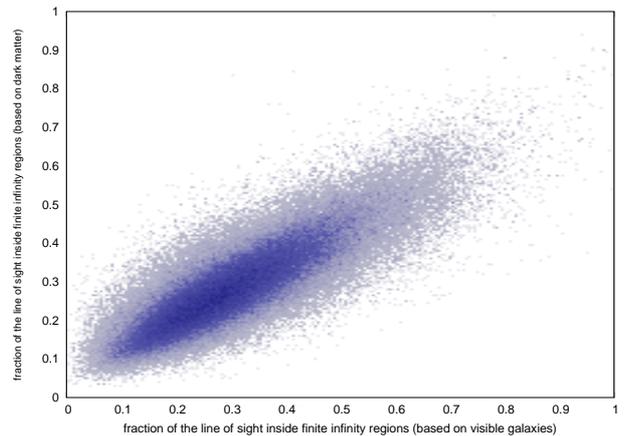}\\ 
\caption{Comparison of fraction of the line of sight within finite infinity regions between the complete dark matter catalogues and the mock catalogues for all lines of sight to our target galaxies in the mock catalogues.}
\label{map_inside_fi_model_vs_dark_f}
\end{center}
\end{figure}

A possible issue with the final results are the simplifications/approximations in timescape mock catalogues. Although improvements by \emph{ad hoc} introduction of backreaction in Newtonian N-body simulations have been made \citep{Racz:2017}, no simulation using timescape cosmology has been performed on any scales remotely comparable to the Millennium simulation so far. Therefore, it was necessary to extrapolate from simulations using another cosmology. To gain the best compatibility with $\Lambda$-CDM models, which were used for comparison, the Millennium simulation \citep{millennium} was used as a foundation for the extrapolated timescape cosmology model. By doing so, it was assumed that the later snapshots of the Millennium simulation are a reasonably good representation of the matter distribution in the late Universe. By adding the targeted observational feature of timescape cosmology into the simulated data, we created a timescape-cosmology inspired deformed $\Lambda$-CDM models. This approximated timescape cosmology model suffers not only from our assumptions put into its derivation from the Millennium simulation, but also from all issues affecting the Millennium simulation itself.

A significant issue with the data from the Millennium simulation is the dearth of rich clusters in the simulation compared to the observational data, which we mentioned several times in this paper. It has to be considered as a systematic bias affecting both ($\Lambda$-CDM and differential expansion) sets of mock catalogues in the same way. Aside from using other numerical simulations, which was not feasible due to large volume required and the implemented semi-analytical models to construct the mock catalogues for this test, there was no way to properly correct for this bias. 

The next issue was the fact that all the FoF groups detected in the Millennium simulation together contain less than half the mass/particles used in the simulation. Using the full particle information from the smaller \emph{millimil} run \citep{millennium,Mill_data}, $\sim 80 \%$ of the particles were found to be within the finite infinity regions around the FoF groups, if their masses/sizes were rescaled accordingly \citep{Saulder:2016}. This knowledge was used to extrapolate the mass of the groups and cluster. The remaining $\sim 20 \%$ of the missing mass was included by simply rescaling the finite infinity regions accordingly. Due to the relatively large scatter between the different mock catalogues of the same cosmology (see Section \ref{sec_res}, we found this method to be slightly better suited for our test than the alternative (weakening the slope of the effect of timescape cosmology accordingly, as discussed in \citet{Saulder:2016}). Thereby, the different mock catalogues became better comparable.

When reconstructing the finite infinity regions from observational data, we had to consider the effect of possible areas that are actually inside finite infinity regions, but not detected in the observational data. There are undetected finite infinite regions, because they only host galaxies too faint to register in SDSS. This is a problem that increases with distance. Additionally, there are uncertainties in the sizes of finite infinity arising from our mass estimates, which are between 15 and 20$\%$ compared to the particle data from the smaller \emph{millimil} run \citep{Saulder:2016}. However, since we had simulated data containing the full dark matter information, we could derive the magnitude of these biases. Because the fraction of the line of sight within finite infinity regions is a crucial parameter for our analysis, we focused on the impact of these biases on this parameter. As illustrated in Figure \ref{map_inside_fi_model_vs_dark_f}, we did not only find a scatter of $\sim15\%$ between the reconstructed values from the mock catalogues and the values derived from the dark matter data, but also a systematic shift. We considered this shift in the calculation of the average fraction of the line-of-slights and thereby corrected for it when calibrating our differential expansion models.

Another issue, which requires consideration when using the Millennium simulation is the no longer up-to-date cosmology used for it. Over the past decade, our knowledge of the cosmological parameters improved, causing a notable difference between the cosmological parameters of the simulation and the cosmological parameters obtained by the most recent CMB observations \citep{Planck,Planck_cosmo}. There were reruns \citep{Guo:2013} using a more present-day set of cosmological parameters (from WMAP7 \citep{WMAP_7}) and also rescaled versions of the original run \citep{Guo:2013}. However, the public database of the reruns does not contain the FoF group catalogue, which was essential for the calibration of the group finder (Paper IV) and the finite infinity regions derived from its results. Furthermore, the rescaled versions yield deviant (incompatible with the observations) number densities for the galaxies, which were inserted into the dark matter halos using semi-analytical models \citep{DeLucia:2006}. For these reasons, the original Millennium run was used for our project. The cosmological parameters of the Millennium simulation were used consistently in the final results and every calibration leading up to them. Despite all these issues, the Millennium simulation is the best available model, which has a sufficiently large volume and all the additional data (semi-analytical galaxy models with SDSS magnitudes, FoF catalogue) that was need for our cosmological test. 

The next question, which has to be addressed, is if artificially introducing the different expansion rates of voids and walls for the timescape model, which were obtained by fitting the timescape model \citep{Wiltshire:2007,Leith:2008,Duley:2013,Nazer:2015} to observed data of supernovae Type Ia \citep{Riess:2007,Kessler:2009}, CMB \citep{WMAP_pre,WMAP3,Planck} and Baryonic acoustic oscillations \citep{Cole:2005,Eisenstein:2005}, into the data of the Millennium simulation was justified. To this end, it is important to remember how the Hubble expansion in general is considered in n-body simulations. The metric expansion of space is a general relativistic effect, which has to be incorporated in an otherwise fully Newtonian simulation. The expansion of the Universe was handled by using co-moving coordinates, which also required to add an additional term (similar to a drag force) to the Newtonian gravity. Consequently, the Hubble expansion in the $\Lambda$-CDM simulation was already introduced as an artificial feature rather than an intrinsic property. Additionally, a present-day timescape cosmology Universe would appear more evolved (which could fix the dearth of rich cluster problem in the Millennium simulation to a degree) than a present-day $\Lambda$-CDM cosmology Universe. Hence, one would ideally prefer snapshots later than present-day (in $\Lambda$-CDM cosmology), but there are no snapshots available beyond present-day, because the simulation was terminated at that point. So again, our differential expansion model is limited by the available data. Because the present-day matter distribution (including peculiar velocities) was the only feature drawn from the simulation, calculating slightly different redshift values from this data in the last snapshots, where structure formation was slow, was therefore a justified approximation.  Naturally, a complete cosmological simulation using timescape cosmology would be preferable and more self-consistent, but such a thing does not exist. Using the Millennium simulation instead was the best practicable option.
 
\subsection{Fundamental plane calibrations}
The fundamental plane of elliptical galaxies was used as a redshift independent distance indicator for our cosmological test. It was calibrated and discussed in detail in \citet{Saulder:2013} and further improvements were made in \citet{Saulder:2015a}. In \citet{Saulder:2016}, a systematic variation of the residual distance dependence of the ratio between the fundamental plane distance and redshift distance depending on the multiplicity of galaxy groups was detected. The fundamental plane distances, using the calibrations of \citet{Saulder:2015a}, have a small residual distance/redshift dependence. However, this bias seems to strongly increase with the richness of the groups. This effect was not detected in the mock catalogues, where the fundamental plane distances were not derived from the galaxies' parameters, but just from their co-moving distances and a statistical scatter (also considering the residual redshift dependence of the fit). Hence, it is fair to assume that this is an environmental effect on the fundamental plane parameters. This would be in agreement with the findings of \citet{Joachimi:2015}. To test how the residual redshift dependence (not the dependence on the richness of the groups) influences the results of the cosmological test, a set of mock catalogues was created, which did not include the residual redshift dependence in the mock fundamental plane distance. The cosmological test was performed using this alternative set of mock catalogues instead of the previously used ones. A comparison showed that no significant differences could be found between the alternative results and the results presented in the Section \ref{sec_res}. We conclude that the variations in the fundamental plane residuals do not have any significant impact on the outcome of the cosmological test presented here. 

\subsection{Normalization}
Normalization and rescaling can be handled easily for our data, because we used relative values for the individual Hubble parameters. This was important, because the absolute Hubble parameters in $\Lambda$-CDM cosmology and timescape cosmology are significantly differently. The Hubble parameter used in the Millennium simulation was 73 km s$^{-1}$ Mpc$^{-1}$ and the best estimates of this value today is (67.8 $\pm$ 0.9) km s$^{-1}$ Mpc$^{-1}$ based on CMB data \citep{Planck,Planck_cosmo}. In contrast to this, the values predicted by timescape cosmology are comparably low at 61.7 $\pm$ 3.0 km s$^{-1}$ Mpc$^{-1}$, being a non-linear average of 50.1 $\pm$ 1.7 km s$^{-1}$ Mpc$^{-1}$ for walls and an apparent maximum of $75.2_{-2.6}^{+2.0}$ km s$^{-1}$ Mpc$^{-1}$ across voids \citep{Duley:2013,Wiltshire:2013}. These values of the Hubble parameter are notable lower than the one from $\Lambda$-CDM cosmology due to the higher age (even in walls) of the Universe in timescape cosmology. To ensure a comparability of the test data with both cosmologies, we used relative values for the (individual) Hubble parameters. We applied rescaling on all of our datasets, so that all had the same average values for the individual Hubble parameters. 

\subsection{Distance/redshift limit}
\begin{figure}[ht]
\begin{center}
\includegraphics[width=0.45\textwidth]{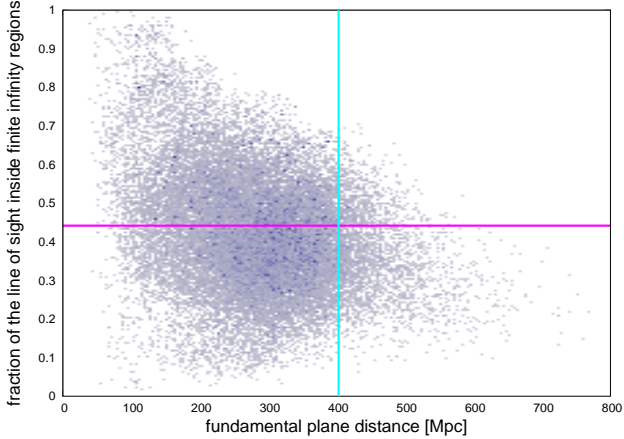}\\ 
\caption{Distance dependence of the full observations data. Magenta dotted line: maximum distance for the selected sample. Magenta line: average fraction of the line of sight within finite infinity regions of the observational data; cyan line: distance limit for our selected sample.}
\label{fi_dist}
\end{center}
\end{figure}
For this test, we limited our dataset to a redshift of 0.1 for the groups hosting early-type galaxies and to 0.11 for the finite infinity regions. Including more distant galaxy groups would not improve the quality of the test, because at greater distances the fraction of the line of sight to one group approaches the average wall fraction of the cosmological model. As illustrated in Figure \ref{fi_dist}, the range of fractions of finite infinity regions in the line of sight shrinks with increasing distances. The highest and lowest fractions, which are the most important ones for our cosmological test, become rarer at larger distances. Therefore, deeper samples would not be able to further improve the quality of test. When we created the selected sample, we also added, aside from other constraints, a cut-off at 402.8 Mpc for the co-moving fundamental plane distances, which corresponds to a redshift of 0.1 using the cosmology of the Millennium simulation. Thereby, we ensure a symmetric cut-off in redshift and actual distance. We also experimented with lower cut-off distances/redshifts, but did not find any notable improvement, because the number counts in our mock-catalogues and observational data shrank as well.

\subsection{Sky coverage}
\begin{figure*}[ht]
\begin{center}
\includegraphics[width=0.90\textwidth]{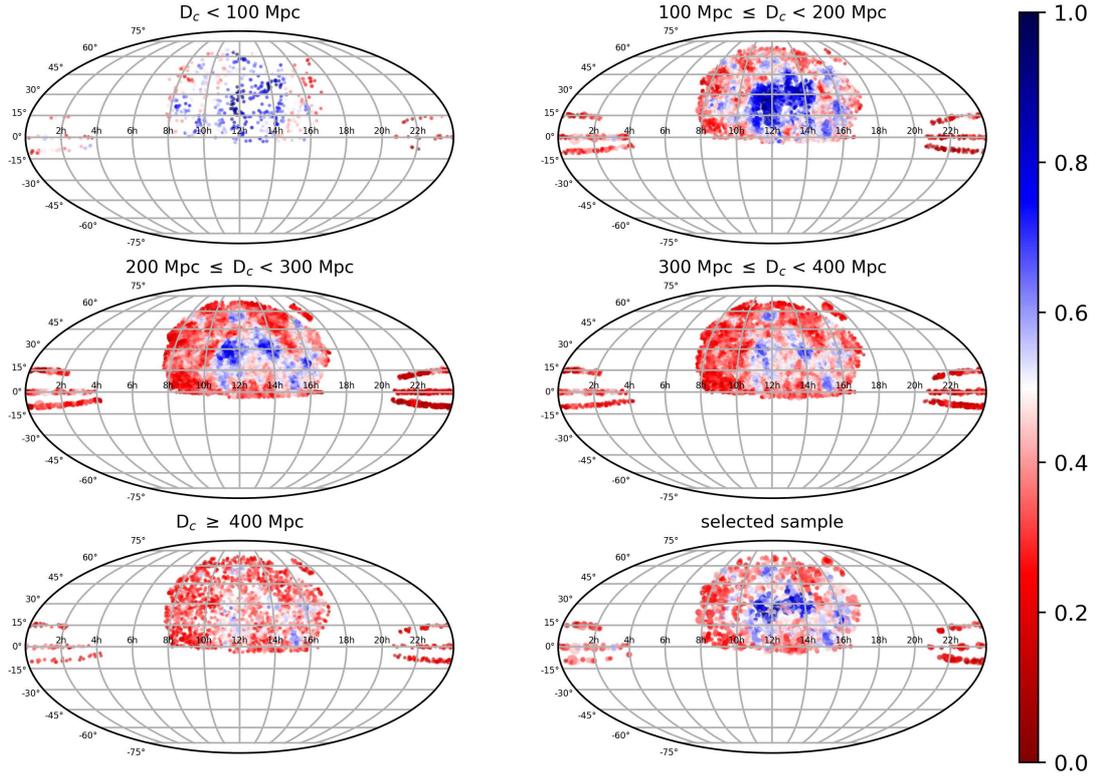}\\ 
\caption{Distribution of the fraction of the line of sight within finite-infinity regions on the sky for different distance slices. The first five of the six panels show the full sample for different slices in co-moving distance, while the last panel shows the selected sample for its entire distance range. The area covered by the (overlapping and partially transparent) circles in the individual depends on the richness of the group to which the fraction of the line of sight within finite-infinity regions was calculated. The colour coding provides the value of the fraction of the line of sight within finite-infinity regions for the individual targets.}
\label{sky_distribution}
\end{center}
\end{figure*}
Large-scale surveys that contain spectroscopic data with sufficient quality to derive central velocity dispersions are still rare. The spectroscopic sky-coverage of SDSS is only about one quarter of the sky \citep{SDSS_DR12}, mostly limited to the northern hemisphere. The only other large-scale ($\sim 10 000$ targets) survey with sufficient data to derive the fundamental plane parameters (especially the central velocity dispersions) is the 6dF Galaxy Survey \citep{6dFGS_first,6dFGS_final}, which covers most of the southern sky. Fundamental plane data was provided for nearly 10 000 galaxies by \citet{Magoulas:2012,Campbell:2014}. There are a couple of reasons, why we did not include that data in our study: the 6dF Galaxy Survey is shallower than SDSS, which would make modelling the finite infinity regions from the observational data even more difficult; it relies on old 2MASS photometry \citep{2MASS}; manual target selection for early-type galaxies, which is difficult to reproduce. For all these reasons, it would be very difficult to combine the 6dF Galaxy Survey data with SDSS data to increase the sky coverage without gaining significant systematic biases between the two hemispheres. The currently ongoing Taipan survey \citep{Taipan}, which in its design is much closer to SDSS and also aims to study peculiar motions using the fundamental plane, might be an interesting dataset to increase the sky coverage of our test in the future. In the very long run, the planned extension of SDSS to a full sky survey (SDSS-V white paper \citep{SDSS_future}) might be the most self-consistent way to improve the cosmological test presented in this paper, thereby eliminating possible systematic biases from only having data from one hemisphere.

The area of the sky covered by the SDSS spectroscopic sample contains a couple of notable features as illustrated in Figure \ref{sky_distribution}. The Virgo-cluster area is near the centre of the SDSS footprint and yields the high values of fractions of the line of sight within finite infinity regions in the central area of the nearest distance slice (<100 Mpc) presented in Figure \ref{sky_distribution}. In the next distance slices the effect of the Coma supercluster and the Leo supercluster\footnote{Based on the coordinates from the NASA Extragalactic Database (NED)} is clearly visible. In the most distant slice in Figure \ref{sky_distribution}, the spatial distribution of the fractions of the line of sight within finite infinity regions has become mostly homogeneous compared to the other slices, which is in agreement with the distance distribution shown Figure \ref{fi_dist}. Our dataset does not cover the local void, but also misses out on other great superstructures like the Great Attractor region. While the SDSS region overall is not outstandingly over- nor under-dense, the fact that it covers the Virgo-cluster, which has significant influence on the local peculiar velocity field, has to be acknowledged as a possible source of systematic errors.

\subsection{Statistical analysis}
The results of the statistical tests performed in Section \ref{sec_res} tend to favour the $\Lambda$-CDM model over the differential expansion model (timescape-cosmology inspired deformed $\Lambda$-CDM model). One issue we encountered during the analysis was that the variations (see Figures \ref{result_scatter_lcdm} and \ref{result_scatter_ts}) between the different mock catalogues using the same cosmology were much larger than we initially expected. These variations were largely caused by two effects: the uncertainties in the fundamental plane distances and the distribution of the matter in the local Universe. There is little that can be done about the latter, but one can reduce the scatter caused by the fundamental plane by selecting galaxy groups that host more than one early-type galaxy. As illustrated in Figures \ref{result_scatter_lcdm_lim} and \ref{result_scatter_ts_lim}, the variations in the mock catalogues were reduced slightly by limiting the galaxy groups used in the analysis to only those that had at least three members. Due the dearth of rich groups in the Millennium simulation, the number of groups in the mock catalogues shrank disproportionally, which partially attenuated the reduction of the scatter. Concerning the variations in the distribution of the matter between the different mock catalogues, we decided to look for the mock-catalogues, which had values for the average fraction of the line of sight within finite infinity regions closest to the ones derived from observational data of 0.47 (for the selected sample). We found that for the selected sample the mock catalogues that had the highest average fraction of the line of sight within finite infinity regions between 0.4 and 0.5 yielded the flattest slopes. Apparently, the mock catalogues that contain a similar richness of finite infinity regions as the observational data were also the best fits (see Figure \ref{result_scatter_lcdm_lim}) to observational data in the case of $\Lambda$-CDM cosmology. They also perform best at the KS- and $\mathcal{X}^{2}$ tests.  

In our comprehensive analysis, we fitted a weighted linear regression on the distribution of relative individual Hubble parameters versus the fractions of the line of sight inside finite infinity regions. We compared the results of the mock-catalogues to the observational data in two different rest-frames (CMB and LG). For full sample, we found that the observational data agrees well with expectations of $\Lambda$-CDM cosmology, and the differential expansion model is already a deviates by several standard deviations (see Figures \ref{map_obs_cmb} and \ref{map_obs_lg}). However, the LG rest frames yields slightly better results and an inspection of the slopes of the individual mock catalogues (see Figure \ref{result_scatter_ts}) show still some compatibility between our differential expansion model and the observations. The selected sample (Figures \ref{map_obs_cmb_selected} and \ref{map_obs_lg_selected}) provides some striking results. The observational data yields a slope that can almost be considered too flat for $\Lambda$-CDM cosmology (although considering our deliberations from the previous paragraph, the obtained values are compatible with $\Lambda$-CDM cosmology given the variance in the mock catalogues) and is clearly incompatible with our differential expansion model. A comprehensive list of our results and probabilities are provided in Table \ref{results_k}. We only used the slope of the linear regression to calculate the probabilities and likelihoods, because the offset is sensitive to normalization and is not the defining feature of our test. 

The distribution of the parameters showed, despite the large scatter, indications of non-linearities, in particular disproportionally high values of relative individual Hubble parameters at low fractions of finite infinity regions in the line of sight. Hence, we used an alternative approach and calculated the average relative individual Hubble parameters in a set of overlapping bins for all combined mock catalogues (see Figures \ref{result_bin_lcdm_lim} and \ref{result_bin_ts_lim}). As illustrated in Figures \ref{finalresults_bins_lim_cmb} and \ref{finalresults_bins_lim_lg}, the observations lie within the error bars of both models when considering each foreground density bin individually. To complete our analysis of the binned data, we fitted a linear regressions on the binned data and studied the average slopes obtained for the different cosmology. The results (see Figures \ref{binfit_selected_cmb} and \ref{binfit_selected_lg} and Table \ref{results_binfitlin}) show again a preference for $\Lambda$-CDM cosmology over the differential expansion model, but by far not as striking as in case of the direct linear regressions. The slope of the approximated timescape model deviates only by $\sim 2-\sigma$ in the case of the preferred (from a timescape point of view) Local Group rest frame. Furthermore, when looking at the bins individually most actually slightly prefer the differential expansion model, albeit due to the huge error bars at an extremely low significance level.

In an alternative approach to classic fitting and binning, we also performed a two-way two-dimensional KS test on our dataset. The test told us that most pairs (corresponding $\Lambda$-CDM cosmology and differential expansion model) of mock catalogues were not compatible with the observational data, which caused extremely low probability values for both cosmological theories. However, one has to consider that for such large samples the KS test's output probability is very sensitive to any slight shifts in data distribution, which may partially explain the low probabilities. For all pairs of mock catalogues, the probabilities values of $\Lambda$-CDM cosmology are higher than of the differential expansion model (very often by huge factors). It should be pointed out that the KS test and the probability values (P values) in general do not proof models, but just tell how likely two datasets, in this case the observational data and one of the mock catalogues, originate from the same distribution. Additionally, we performed an $\mathcal{X}^{2}$ test on our data. Again the significance levels are the main problem, since the observational data does not provide a great fit for either simulated dataset. However, for the vast majority of the mock-catalogues pairs $\Lambda$-CDM cosmology is preferred. In the case of the mock catalogues with average fractions of the line of sight within finite infinity regions comparable to the one from observational data, the likelihood for the $\Lambda$-CDM is even more than $50\%$ higher than the one for our differential expansion model.

\section{Summary and conclusions}
We designed and executed a cosmological test in which we compared the measured expansion rates of voids and walls to model predictions from both $\Lambda$-CDM cosmology and the inhomogeneous timescape cosmology \citep{Wiltshire:2007}. To this end, we used observational data based on SDSS \citep{SDSS_DR12} and 2MRS \citep{2MRS} as well as simulated data taken from the Millennium simulation \citep{millennium}. Our test measured the relative Hubble parameters of galaxy groups obtained using the fundamental plane of early-type galaxies, and correlated it with the matter distribution, the so-called finite infinity regions, in the lines of sight to these galaxies. To obtain these parameters, we made full use of our previous work building up to this test: the intitial design of \citet{Saulder:2012}, the fundamental plane calibration of \citet{Saulder:2013} with their improvements in \citet{Saulder:2015a}, and the model of the matter distribution in the local Universe of \citet{Saulder:2016}. Executing our test, we compared the observed variations to the expectations of both $\Lambda$-CDM cosmology and an approximation of timescape cosmology, which were derived from mock catalogues based on the Millennium simulation. We performed various statistical analyses on the results, which generally showed a preference for $\Lambda$-CDM cosmology when compared to our differential expansion model, which we used to approximate a phenomenological prediction of timescape cosmology.

From a $\Lambda$-CDM cosmology point of view, the outcome of our test seems relatively clear and straightforward. The results of our analyses agree well with the predictions of $\Lambda$-CDM cosmology and the only notable outlier is in the opposite direction of the effect proposed by timescape cosmology and other inhomogeneous models. When considering the scatter and the comparability of the individual mock catalogues, even the flat slope in Figure \ref{map_obs_cmb_selected} agrees with the likely best-suited mock catalogue (see Figure \ref{result_scatter_lcdm_lim}). All tests concluded that $\Lambda$-CDM cosmology is more likely to fit the observations than the approximated timescape model and there are no signs of any significant deviations indicating strong backreaction effects in the observational data. However, due to the scatter and systematic uncertainties affecting our current data, we cannot explore other backreaction models \citep{Clarkson:2012,Umeh:2014a,Umeh:2014b,Clarkson:2014} that affect the cosmological parameters on a percentage level.

From a timescape cosmology point of view, the results are much more difficult to interpret. Timescape cosmology is a conceptually interesting theory, which could explain dark energy in an elegant way by an apparent effect from backreactions caused from observed inhomogeneities (voids and walls). This would would have eliminated the need to expand General Relativity by an additional parameter, the cosmological constant. However, our cosmological test shows that the differential expansion model, which we used here to approximated timescape cosmology, yields notably worse fits to the observational data than $\Lambda$-CDM cosmology. The approximation, which we used to derive mock data for timescape cosmology, was a deformed (introducing the different expansion rates artificially) $\Lambda$-CDM model. We argue that this approach is sufficient to gain at least a rough estimate of how (direction and magnitude) the signal produced by timescape cosmology would manifest itself in observational data. Our model did not consider any potential screening mechanisms or changes in peculiar motion field for an older Universe as expected in timescape cosmology. However, given the notable difference between apparent (for a wall observer, such as we are) expansion rates in voids and in walls in timescape cosmology, we consider them to dominate any other effect on the peculiar motion field. The Local Group rest frame generally provides better fits than the Cosmic Microwave Background rest frame, especially for our approximated timescape model. We consider this very interesting because the rest frame of timescape cosmology is expected to be closer to the LG rest frame (which represents our own finite infinity region) than to the CMB rest frame. Since SDSS is focused on the northern hemisphere and there are no full-sky surveys to draw self-consistent fundamental plane data from, we cannot completely investigate the full impact of different rest frames and systematic biases caused by them. In the end, although noteworthy, the difference between the two rest frames is too small to turn tides in the statistical analysis. Some of the statistical tests do not exclude timescape cosmology. We found that the KS-test and the $\mathcal{X}^{2}$ test provide poor fits to our data in both cosmologies, which can be attributed to various unavoidable flaws in the mock catalogues, as discussed earlier. In both tests, the mock catalogues with matter distributions closest to the observations show the best-fitting values and also the most notable agreements with $\Lambda$-CDM cosmology over timescape cosmology. The analysis of the binned data showed that timescape cosmology is compatible with the error bars of essentially all bins and slightly elevated values (but still within $\Lambda$-CDM error bars) of the relative individual Hubble parameters for predominantly void lines of sight (see Figures \ref{finalresults_bins_lim_cmb} and \ref{finalresults_bins_lim_lg}) can be found in the data. Although the slope fit on the binned data still prefers $\Lambda$-CDM cosmology, the slope of our approximated timescape model is only $2-\sigma$ away from the observational data. Considering the statistical errors between 10$\%$ and 20$\%$ from the distance estimator and the model the matter distribution (finite infinity regions) as well as possible systematics from the specific mock catalogues, one cannot fully exclude timescape cosmology with our current quality of data and models. Nevertheless, our test did not produce any significant evidence in favour of timescape cosmology either.

We conclude that despite some uncertainties in the models, our test yields results in favour of $\Lambda$-CDM cosmology over the differential expansion model. Following the paradigm that extraordinary claims require extraordinary evidence, this evidence in support of timescape cosmology cannot be found in the observational data. The statistical analysis consistently prefers $\Lambda$-CDM cosmology over timescape cosmology. Any signal produced by differential expansion on the magnitude required by timescape cosmology should have yielded a detectable deviation from the $\Lambda$-CDM expectations towards the values derived from our differential expansion model, which is based on an approximated timescape cosmology. We conclude that we cannot find any clear indications of an impact of inhomogeneities on the expansion rates that is sufficiently strong to explain dark energy completely.

To improve the significance of the tests performed in this paper, there are several possible avenues: As already discussed, one would not gain much additional information from deeper surveys. However, with wider survey data, especially with full-sky coverage, we would be able reduce the uncertainties. Furthermore, one could use other distance indicators as well to exclude potential systematic biases from the fundamental plane of early-type galaxies. More advanced numerical simulations to derive mock catalogues could also improve the results of our test, especially a self-consistent simulation using timescape cosmology. Future investigations in this field should focus on finding strong limits for the impact of backreactions caused by large-scale inhomogeneities on cosmological parameters. 

\section*{Acknowledgments}
Funding for SDSS-III has been provided by the Alfred P. Sloan Foundation, the Participating Institutions, the National Science Foundation, and the U.S. Department of Energy Office of Science. The SDSS-III web site is \url{http://www.sdss3.org/}.

SDSS-III is managed by the Astrophysical Research Consortium for the Participating Institutions of the SDSS-III Collaboration including the University of Arizona, the Brazilian Participation Group, Brookhaven National Laboratory, University of Cambridge, Carnegie Mellon University, University of Florida, the French Participation Group, the German Participation Group, Harvard University, the Instituto de Astrofisica de Canarias, the Michigan State/Notre Dame/JINA Participation Group, Johns Hopkins University, Lawrence Berkeley National Laboratory, Max Planck Institute for Astrophysics, Max Planck Institute for Extraterrestrial Physics, New Mexico State University, New York University, Ohio State University, Pennsylvania State University, University of Portsmouth, Princeton University, the Spanish Participation Group, University of Tokyo, University of Utah, Vanderbilt University, University of Virginia, University of Washington, and Yale University. 

This publication makes use of data products from the Two Micron All Sky Survey, which is a joint project of the University of Massachusetts and the Infrared Processing and Analysis Center/California Institute of Technology, funded by the National Aeronautics and Space Administration and the National Science Foundation.

This research has made use of the NASA/IPAC Extragalactic Database (NED) which is operated by the Jet Propulsion Laboratory, California Institute of Technology, under contract with the National Aeronautics and Space Administration. 

We also want to thank our referee David Wiltshire for his very extensive reports that were sufficiently extensive to be published as papers of their own.

\addcontentsline{toc}{section}{References}
\bibliography{paper}\label{bib}

\appendix
\section{Updated finite infinity regions catalogue}
\label{newfiregions}
\begin{table*}[ht]
\begin{center}
\begin{tabular}{cccccccccc}
 RA &  DEC & $D_{c}$ & $c_{x}$ & $c_{y}$ & $c_{z}$ & $\textrm{log}_{10} \left( M_{\textrm{fi}} \right)$ & $\textrm{log}_{10} \left(\Delta M_{\textrm{fi}} \right)$ & $R_{\textrm{fi}}$ & $ \Delta R_{\textrm{fi}}$\\
 $[^\circ]$ & $[^\circ]$ & $[\textrm{Mpc}]$ & $[\textrm{Mpc}]$ & $[\textrm{Mpc}]$ & $[\textrm{Mpc}]$ & [$\textrm{log}_{10} \left( M_{\astrosun} \right)$] & [$\textrm{log}_{10} \left( M_{\astrosun} \right)$] & $[\textrm{Mpc}]$ & $[\textrm{Mpc}]$\\\hline 
-177.7913 & 19.4997  &  16.776  & -15.802  &  -0.609 &    5.600 & 15.59534 &  0.17432 & 21.28614  & 6.06958  \\
  54.5391 &-34.8066  &  20.178  &   9.612  &  13.495 &  -11.518 & 14.51758 &  0.17432 &  9.30770  & 2.65402  \\
  40.9810 & -0.1748  &  16.912  &  12.768  &  11.091 &   -0.052 & 13.00364 &  0.17432 &  2.91203  & 0.73113  \\
  40.1001 & 39.0633  &   9.640  &   5.725  &   4.821 &    6.075 & 12.88081 &  0.17432 &  2.65003  & 0.66535  \\
  62.8739 & 69.8577  &  18.152  &   2.850  &   5.563 &   17.042 & 13.61868 &  0.17432 &  4.66886  & 1.33129  \\
 -20.7329 & 34.4159  &  13.766  &  10.621  &  -4.020 &    7.781 & 12.18013 &  0.17432 &  1.54772  & 0.38859  \\
 -72.5579 &-63.8575  &  18.323  &   2.420  &  -7.702 &  -16.449 & 12.49314 &  0.17432 &  1.96802  & 0.49411  \\
 145.6387 & -3.6991  &  27.922  & -23.001  &  15.726 &   -1.801 & 13.25438 &  0.17432 &  3.53001  & 0.88629  \\
  65.0666 &-56.5689  &  20.813  &   4.834  &  10.398 &  -17.370 & 13.87016 &  0.17432 &  5.66289  & 1.61473  \\
 146.2288 &-31.3025  &  20.639  & -14.659  &   9.803 &  -10.723 & 13.06682 &  0.17432 &  3.05671  & 0.87160

\end{tabular}
\end{center}
\caption{Parameters of the first ten groups as they appear in our finite infinite regions catalogue in the Local Group rest frame. Columns 1 and 2: equatorial coordinates (right ascension and declination); column 3: co-moving distance; columns 4 to 6: the Cartesian co-moving coordinates of the centre of the finite infinity region; column 7: the mass within the finite infinity region; column 8: the uncertainty for mass within the finite infinity regions; column 9: the radius of the finite infinity region; and column 10: the uncertainty in the radius of the finite infinity region.}
\label{sample_cat_fi_lg}
\end{table*} 

\begin{table*}[ht]
\begin{center}
\begin{tabular}{cccccccccc}
 RA &  DEC & $D_{c}$ & $c_{x}$ & $c_{y}$ & $c_{z}$ & $\textrm{log}_{10} \left( M_{\textrm{fi}} \right)$ & $\textrm{log}_{10} \left(\Delta M_{\textrm{fi}} \right)$ & $R_{\textrm{fi}}$ & $ \Delta R_{\textrm{fi}}$\\
 $[^\circ]$ & $[^\circ]$ & $[\textrm{Mpc}]$ & $[\textrm{Mpc}]$ & $[\textrm{Mpc}]$ & $[\textrm{Mpc}]$ & [$\textrm{log}_{10} \left( M_{\astrosun} \right)$] & [$\textrm{log}_{10} \left( M_{\astrosun} \right)$] & $[\textrm{Mpc}]$ & $[\textrm{Mpc}]$\\\hline 
 148.9685  &  69.6797  &  3.058  &  -0.910  &  0.547  &   2.868  & 12.67324 & 0.17432 &  2.25977  & 0.56758 \\   
-158.5522  & -42.7874  &  9.657  &  -6.596  & -2.591  &  -6.560  & 13.42277 & 0.17432 &  4.01703  & 1.00895 \\     
-163.6363  & -49.4679  & 11.034  &  -6.880  & -2.020  &  -8.386  & 12.99607 & 0.17432 &  2.89516  & 0.72717 \\   
 176.6011  &  51.6091  & 14.384  &  -8.917  &  0.530  &  11.274  & 14.44156 & 0.17432 &  8.78018  & 2.50360 \\
-175.7590  &  13.6746  & 20.2072 & -19.581  & -1.452  &   4.777  & 15.52615 & 0.17432 & 20.18515  & 5.75564 \\
  54.1144  & -34.4517  & 18.0024 &   8.702  & 12.027  & -10.184  & 14.52948 & 0.17432 &  9.39313  & 2.67838 \\
  40.9097  &  -0.1378  & 13.5450 &  10.236  &  8.870  &  -0.033  & 13.00691 & 0.17432 &  2.91935  & 0.73325 \\
  39.8015  &  38.7707  &  5.3489 &   3.204  &  2.670  &   3.349  & 12.90131 & 0.17432 &  2.69208  & 0.76763 \\     
  62.9295  &  69.9077  & 15.2306 &   2.381  &  4.659  &  14.304  & 13.62085 & 0.17432 &  4.67663  & 1.33350 \\ 
 -20.7329  &  34.4159  &  6.5194 &   5.0230 & -1.904  &   3.685  & 12.17963 & 0.17432 &  1.54712  & 0.38859   
\end{tabular}
\end{center}
\caption{Parameters of the first ten groups as they appear in our finite infinite regions catalogue in the CMB rest frame. Columns are the saame as in Table \ref{sample_cat_fi_lg} }
\label{sample_cat_fi_cmb}
\end{table*} 

Since the catalogue of finite infinity regions in \citet{Saulder:2016} used slightly outdated values \citep{Wiltshire:2007,Leith:2008} for the the cosmological parameters of timescape cosmology, we provide an updated catalogue, which uses the latest best fit values from \citet{Duley:2013}. We also added additional columns to our catalogue to provide complementary information on the position (right ascension and declination) of the finite infinity regions on the sky. Furthermore, we provide the catalogue of finite infinity regions calculated for two different rest systems (CMB for $\Lambda$-CDM cosmology and LG for timescape cosmology). In Tables \ref{sample_cat_fi_lg} and \ref{sample_cat_fi_cmb}, we present a first 10 rows of the finite infinity regions catalogue in the Local Group rest frame and the CMB rest frame as an example. The full data is made available online on VizieR.

\end{document}